\documentclass[aps,pra,reprint,longbibliography,superscriptaddress]{revtex4-1}
\usepackage{hyperref}  
\usepackage{graphicx}  
\usepackage{dcolumn}   
\usepackage{bm}        
\usepackage{amssymb}   
\usepackage{verbatim}  
\usepackage{epstopdf}
\usepackage{color}
\usepackage{soul}
\usepackage{amsmath}
\usepackage{subfigure}
\newcommand{\commentout}[1]{{}}
\usepackage{latexsym}
\usepackage{mathrsfs}
\usepackage{bbold}

\newcommand{\bea}{\begin{eqnarray}}
\newcommand{\eea}{\end{eqnarray}}
\newcommand{\beq}{\begin{equation}}
\newcommand{\eeq}{\end{equation}}

\newcommand{\half}{\hbox{$\frac{1}{2}$}}
\newcommand{\threehalves}{\hbox{$\frac{3}{2}$}}
\newcommand{\eq}[1]{(\ref{#1})}

\newcommand{\br}{{\bf r}}

\newcommand{\bk}{{\bf k}}

\newcommand{\bE}{{\bf E}}

\newcommand{\bd}{{\bf d}}

\newcommand{\G}{{\sf G}}

\begin{document}
\title{Light propagation in systems involving two-dimensional atomic lattices}
\author{Juha Javanainen}
\affiliation{Department of Physics, University of Connecticut, Storrs, Connecticut 06269-3046}
\author{Renuka Rajapakse}
\affiliation{Department of Physics, University of Massachusetts, Dartmouth 02747}
\affiliation{Department of Physics, University of Connecticut, Storrs, Connecticut 06269-3046}
\begin{abstract}
We study the optical response of a 2D square lattice of atoms using classical electrodynamics. Due to dipole-dipole interactions, the lattice atoms polarize as if the lattice were an atom with up to three resonance frequencies, with cooperatively shifted resonances and altered transition linewidths. We show that when the distance between two 2D lattices is large enough and Bragg reflections are absent, the lattices interact  among themselves as if they radiated a plane wave whose amplitude is in accordance with the radiation from a dipole moment continuously distributed in the lattice plane. We employ these results to study light propagation in stacks of 2D lattices, drawing on simple qualitative pictures of the response of a 2D lattice and light propagation in 1D waveguides. We show that a stack of 2D lattices may emulate regularly spaced atoms in a lossless 1D waveguide, and argue that in a suitable geometry the resonance shifts characteristic of 1D and 2D lattice structures may completely cancel to eliminate density dependent resonance shifts of atoms bound to a 3D lattice. A generalization to the case of anisotropic polarizability, such as in the presence of a magnetic field, reveals light frequencies induced by the magnetic field for which the lattice is either completely transparent, or completely opaque.
\end{abstract}
\maketitle
\section{Introduction}
Advances in experimental~\cite{Keaveney2012,Pellegrino2014a,wilkowski2,Ye2016,Guerin_subr16,Jennewein_trans,Roof16,COR17} and numerical~\cite{Javanainen1999a,CHO12,Bienaime2013,Javanainen2014a,Ye2016,Sutherland1D,ZHU16,JAV17} techniques have revitalized classical electrodynamics of material samples as a topic of frontline research.  The idea is that in cold, dense atomic samples dipole-dipole interactions between the atoms mediated by light may lead to cooperative behavior of light-matter systems, and even result in a strongly correlated sample. It transpires that the standard classical electrodynamics of polarizable media~\cite{Jackson,BOR99} is a mean-field approximation, and as such could be inaccurate in cold dense samples~\cite{Javanainen2014a,JavanainenMFT}.

Aside from 3D gaseous samples, cooperation of atoms in a 2D lattice has attracted recent interest~\cite{Jenkins_2d_lat,Bettles_prl16,YOO16,Facchinetti,FacchinettiLong,SHA17,PEROR17,PER17,ASE17,BET17,YOO18}. In fact, studies of metamaterials with 2D arrays of dipoles that mix electric and magnetic dipole moments have made similar theoretical points earlier~\cite{CAIT}, and seem to be ahed of atomic lattices even in the experiments~\cite{JEN17}, but we focus on the simpler case of ordinary atoms. For the present purposes the observation that a lattice with atoms may reflect back all of the incident light~\cite{Jenkins_2d_lat,Bettles_prl16,SHA17} is of particular interest. Going down in the number of dimensions, effectively 1D waveguides holding atoms are also a subject of long-standing theoretical and experimental activity~\cite{DEU95,KIE04,Rauschenbeutel,TIE14,kimblenaturecom,kimblemanybody,kimblesuper,SOR16,COR16,RUO16,RUO17}. Light-induced correlations in 1D~\cite{RUO16,RUO17} has been one particular focus. We will use these results to investigate light propagation in 3D structures formed by stacking 2D lattices.

Given the recent activity in the 2D systems and the overlap of our paper with the existing literature, we open with a broad discussion of how the tenor of the present paper is different. First, as far as the 2D lattices are concerned, we investigate their response and light propagation directly, and mostly exactly. We do not engage in numerical simulations of necessarily finite 2D lattices, nor do we compute cooperative light-atom modes in either finite or infinite 2D lattices. Second, our approach is unapologetically classical electrodynamics. We even describe Zeeman splitting of the atomic levels by using an anisotropic polarizability for the atom. Finally, as a key technical item, in infinite lattices one encounters infinite sums that do not converge absolutely, and present numerical problems that have been addressed in various ways in the past~\cite{SHA17,PER17,BET17}. Here we apply the usual exponential convergence factor, and carry out  the sums in real space using numerical-analysis techniques to accelerate the convergence.

While the correspondence  to the layout of the present paper is not one-to-one, our narrative runs as follows:
We start with one 2D square lattice of atoms. In this case the response of the lattice to light can be studied almost analytically, except for certain infinite sums~\cite{YOO16}. The values of the sums notwithstanding, thanks to the Lorentzian form of the response of an individual atom, a lattice responds to the incident light like an atom with modified resonance frequency and linewidth; or in the case of non-normal incidence of light, like an atom with up to three resonance frequencies.

Since stacking of the 2D lattices is a recurring topic here, we study the transfer of radiation from one lattice to another. While also presenting explicit counterexamples, we find that a simple approximation whereby the lattice radiates as if the atomic dipoles were smeared continuously across a plane is often a good description of the radiation from a 2D lattice.

As a side effect of our studies of light propagation between 2D lattices, we revisit the case of complete reflection from a 2D lattice. Cooperation between the atoms is naturally inseparable from the rest of the physics of the system, but the cooperation per se is not responsible for total reflection: The combination of a single Lorentzian resonance and energy conservation suffices for total reflection. Bragg scattering of light from the 2D lattice may lead to seeming violation of energy conservation, so we also discuss Bragg scattering.

We proceed to combine considerations from different dimensionalities. We point out that a stack of 2D lattices can emulate a perfect 1D waveguide with no losses of light. In another example, by viewing a 3D lattice as as stack of 2D lattices, we show how cooperative shifts in 2D and shifts characteristic of 1D waveguides may combine to eliminate completely the density dependent resonance shifts analogous to collisional shifts. Such shifts occur~\cite{CHA04} even if each atom is bound to a lattice site.

In our final example, basically an alternative take of Refs.~\cite{Facchinetti,FacchinettiLong}, we add to our approach a magnetic field and the attendant anisotropic response of the atoms to the driving field. We demonstrate with numerical and analytical arguments that the magnetic field may induce both perfect reflection and perfect transmission of light through a 2D lattice.

A retrospective discussion in Sec.~\ref{DISCUSSION} closes the present paper.

\section{Setup}
\subsection{Electrodynamics primer}
We take the light to be classical. This can be rigorously justified for an atom with a $J=0\rightarrow J'=1$ transition in the limit of asymptotically low light intensity, and when photon recoil effects are negligible~\cite{Ruostekoski1997a}. The latter is the case if the lattice binds the atoms to regions small compared to the wavelength of light. Of course, the confinement is never perfect, and even if nothing else, the atoms are always subject to quantum mechanical zero-point fluctuations. These could be taken into account, say, by direct numerical simulations~\cite{RUO17}, but here we assume fixed lattice positions of the atoms.

We use custom units~\cite{JAV17} chosen in such a way that the numerical values for the wave number of light and for certain natural constants are
\beq
k = c = \hbar = \frac{1}{4\pi\epsilon_0} = 1\,.
\eeq
As usual, time dependent quantities are written in terms of the slowly-varying positive-frequency quantities, factoring out the time dependence at the frequency $\omega$ of the driving light. The dipole propagator $\G$ that gives the electric field at $\br$ from a dipole with the amplitude $\bd$ at $\br'$ as ${\bf E}({\bf r}) =\G({\bf r}-{\bf r}') \,{\bf d}$ is $3\times3$ matrix, a tensor.  In our units it has the cartesian components
\bea
&&\G_{ij}({\bf r}-{\bf r}') =\nonumber\\ &&\hat{\bf e}_i \cdot\left\{
(\hat{\bf n}\times \hat{\bf e}_j)\times\hat{\bf n} + [3\hat{\bf n}(\hat{\bf n}\cdot\hat{\bf e}_j)-\hat{\bf e}_j]\left( \frac{1}{r^2}-\frac{i}{r}\right)\right\}\frac{e^{ir}}{r}\,.\nonumber\\
\label{DIMLESSG}
\eea
Here $\hat{\bf e}_i$ are the cartesian unit vectors, and $r$ and $\hat{\bf n}$ are the distance from the source point to the field point and the unit vector directed from the source point to the field point. This tensor is a function of the difference between the coordinates, and satisfies $\G_{ij}({\bf r}-{\bf r}')=\G_{ji}({\bf r}-{\bf r}')  = \G_{ij}({\bf r}'-{\bf r})$.

The final salient part of the electrodynamics has to do with the polarization of the atoms. We scale the detuning, the difference $\omega-\omega_0$ between the light frequency and the atomic transition frequency, to the HFHM linewidth $\gamma$ of the optical transition, defining
$$\delta =\frac{\omega-\omega_0}{\gamma}.$$
For a $J=0\rightarrow J'=1$ transition in an atom at low light intensity (and also for a classical isotropic charged harmonic oscillator) the induced dipole moment and the electric field are parallel. With the present units and conventions their quotient, the polarizability, reads
\beq
\alpha = -\frac{3}{2(\delta+ i)}\,.
\label{POLARIZABILITY}
\eeq

\subsection{Self-sum}
\label{SELFSUMCOMP}
We study a square lattice. For definite notation, we say that the squares are aligned with the $x$ and $y$ axes,  and $z$ is the direction perpendicular to the lattice. The lattice spacing is denoted by $a$; $a=2\pi$ would be one wavelength. The lattice sites are specified as ${\bf R}_{\bf n}=a {\bf n}=a(n_x \hat{\bf e}_x+n_y \hat{\bf e}_y)$ with integers $n_x$ and $n_y$. We denote $n=|{\bf n}|$;  ${\bf n}=0$ means the lattice site at the origin. Assume identical dipoles $\bd$ at all lattice sites. Ignoring the self-field of the atom at the origin, the field at the origin is
\beq
\bE(0) = \sum_{{\bf n}\ne0} \G({-\bf R}_{\bf n})\bd\,.
\eeq
By lattice translation invariance, this is actually the field at all sites of the lattice.
If the dipole is circularly polarized  in the plane of the atoms, so obviously is the field, and the now-scalars dipole and electric field are related by
\bea
E(0) &=& S(a) d;\\
S(a) &=& \sum_{{\bf n}\ne0} \frac{e^{i \rho} \left(\rho^2-i \rho+1\right)}{2 \rho^3};\quad \rho = na\,.
\label{RAWSELFSUM}
\eea
Since the result holds for both left- and right-circularly polarized quantities, it actually holds for any polarization of the dipole and the electric field in the plane of the lattice.

The issue here is that the sum is not absolutely convergent. We cure this with the usual convergence factor $e^{-\eta \rho}$, and let $\eta \rightarrow0+$ at the end of the calculation. The other problem from the point of view of numerics is the infinite sum. Our task therefore is to carry out numerically the operations
\beq
S(a) = \lim_{\eta=0+}\left\{\lim_{M\rightarrow\infty} \sum_{{\bf n}\ne0}^{|{\bf n}|\le M}
\frac{e^{(i -\eta)\rho} \left[\rho^2-i \rho+1\right]}{2 \rho^3}\right\}.
\label{SELFSUM}
\eeq
We call $S(a)$ the self-sum. We have developed code that computes $S(a)$ to a preset numerical precision. The technical idea is to accelerate the convergence to the limit $\eta\rightarrow0$ using Richardson extrapolation~\cite{NUMRES}. The details are discussed in Appendix~\ref{NUMDET-R}.

It turns out that the expression~\eq{SELFSUM} does not necessarily converge in the limit $\eta~\rightarrow~0$. When lattice spacing is increased, this happens for the first time at $a=2\pi$, when the lattice spacing equals the wavelength of the driving light. The reason is Bragg scattering. For $a=2\pi$ light scattered from all atoms interferes constructively in the $x$ and $y$  directions, which in the limit of an infinitely large lattice produces an infinite field strength. Such in-plane Bragg scattering happens every time there are parallel lines of lattice points with the shortest distance between the lines equal to a wavelength. Elementary solid state physics adapted from three to two dimensions shows that this happens at lattice spacings $a=2\pi\sqrt{m_1^2+m_2^2}$, where $m_1$ and $m_2$ are integers, and not both equal to zero. A comprehensive discussion of Bragg scattering is deferred to Appendix~\ref{BRAGREF}.

\begin{figure}
  \includegraphics[width=0.9\columnwidth]{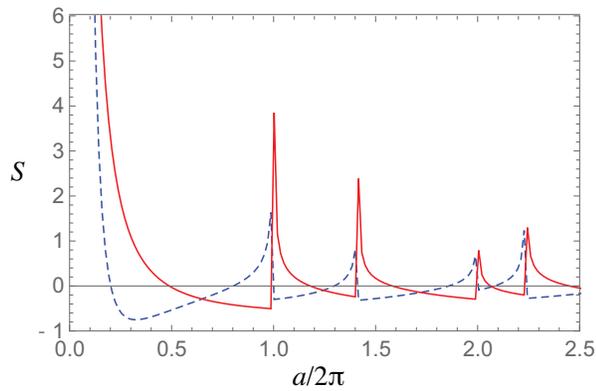}
  \caption{Real (dashed blue line) and imaginary (solid red line) parts of the self-sum $S$ as a function of lattice spacing $a$ scaled to wavelength. There are divergences at the Bragg reflections along the lattice plane, which are truncated in the figure as a result of the finite step between the lattice spacings used for the drawing.
  }
 \label{FS}
\end{figure}

Figure~\ref{FS} shows the self-sum, real (dashed blue line) and imaginary (solid red line) parts as a function of the lattice spacing scaled to $2\pi$ \footnote{There is a functionally equivalent figure in the Supplemental Material of Ref.~\cite{SHA17}}.
The lowest Bragg reflections are expected at $a/2\pi = 1$, 1.41, 2, and 2.24, and those are indeed the exceptional values of the lattice spacing in the figure. The real and imaginary parts are peculiar in that they abruptly jump to infinity when a Bragg-reflection lattice spacing is crossed from above (real part) or below (imaginary part), and diverge smoothly on the other side. Although it is not very obvious from the figure, at small $a$ the real and imaginary parts, respectively, diverge as $1/a^3$ and $1/a^2$. 

\subsection{Transfer sum}
Let us now consider the radiation out of the plane of the lattice. Specifically, imagine an identical second lattice displaced by the distance $z>0$  from the original lattice.  By virtue of lattice translation invariance, the total field from the dipoles of the original lattice is the same at all sites of the second lattice. The field equals $E(0)=T(a,z)d$, where the transfer sum $T$ is
\bea
&&T(z,a) \nonumber\\
&&=\sum_{\bf n}\frac{\left[r^4-i r^3+r^2 \left(z^2+1\right)+3 i r z^2-3 z^2\right] e^{i (r
   -|z|)}}{2 r^5};\nonumber\\
&& r= \sqrt{(na)^2+z^2}\,.
\label{TRFSUM}
\eea

The transfer sum explicitly includes the center site ${\bf n}=0$, whereas in the self-sum the site ${\bf n}=0$ is included only indirectly in the radiative damping rate and the Lamb shift of the atom. Second, the distances between the source and target points $r$ depend on the translation $z$ between the two lattices. Third, the original lattice is expected to radiate a field with a propagation phase $e^{iz}$ toward the increasing coordinate $z$ and $e^{-iz}$ toward the decreasing coordinate. We have explicitly canceled the free-propagation phase from the transfer sum. Because of the $|z|$ in the exponential, the expression~\eq{TRFSUM} is actually valid for both signs of the translation $z$. Finally, the sum~\eq{TRFSUM} is no more absolutely convergent than~\eq{RAWSELFSUM}. This is handled by adding a long-distance convergence factor $e^{-\eta \rho}$ just like in Eq.~\eq{SELFSUM}. The numerical issues are essentially the same as those encountered with  Eq.~\eq{SELFSUM}.

There is also an analytical argument to be made. Namely, for any $z\ne0$ and in the limit $a\rightarrow0$, the terms in the transfer sum vary less and less with the changes of the lattice site index $\bf n$ between neighboring sites, and the sum may be approximated by the integral over $\bf n$. In the integral it is possible to do the limit $\eta\rightarrow0$ analytically as well. All of this makes an interesting exercise in Mathematica, and the result is
\beq
\bar T(a) =\frac{2\pi i}{a^2}\,,
\eeq
independently of the value of $z$. A priori, one expects that this form of the transfer sum is accurate whenever $|z|\gg a$. 

A moment's reflection shows that $\bar T(a)d$ is the electric field that would ensue if there were a continuously distributed dipole moment in the plane, with the surface density $d/a^2$. Deviations of $T(z,a)$ from $\bar T(a)$ can be attributed to the discreteness of the dipoles in the lattice. The lumped nature of the dipoles shows qualitatively in two aspects. First, the transfer sum also has Bragg reflection singularities for the same values of $a$ as the self-sum. Second, in the limit $z\rightarrow0$ with a fixed $a$, the transfer sum obviously converges to the same value as the self-sum, except for the added contribution from the dipole at ${\bf n}=0$. The field of this dipole at the distance $|z|$ diverges with $z\rightarrow0$, so that $T(z,a)$ diverges with $z\rightarrow0$ for any fixed $a>0$.

\begin{figure}
  \centering
  \includegraphics[width=0.7\columnwidth]{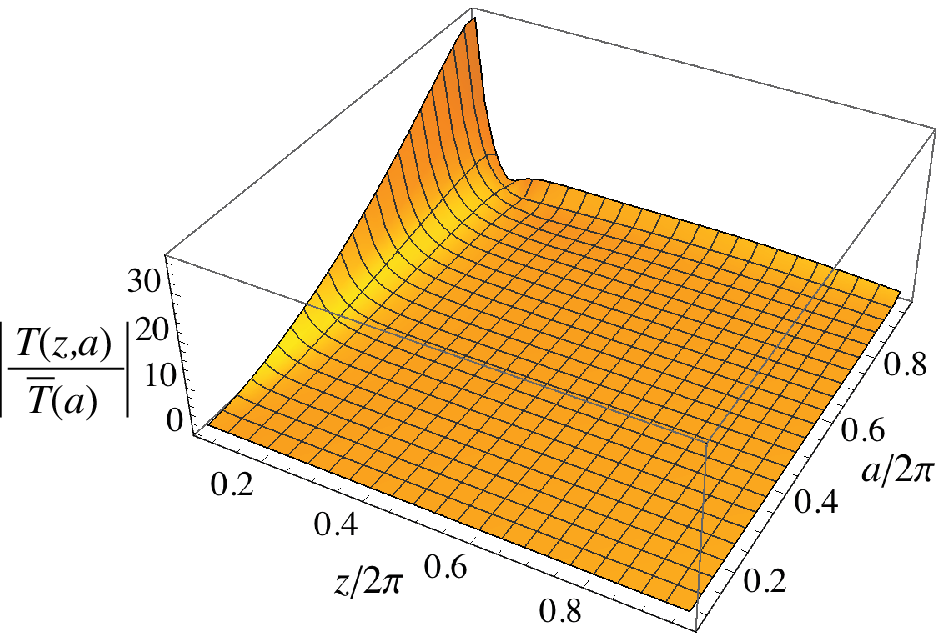}\\\vspace{10pt}\includegraphics[width=0.6\columnwidth]{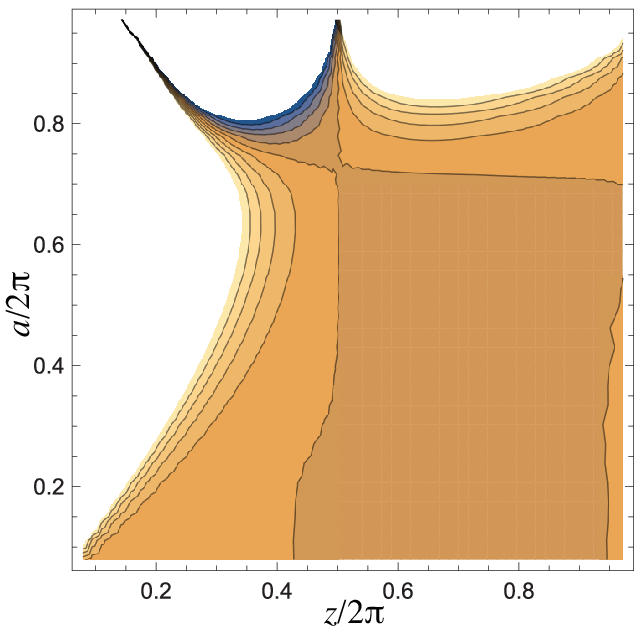}
 \caption{Top: 3D plot of the absolute value of the ratio of the transfer sum $T(z,a)$ to the long-distance limit $\bar T(a)$ as a function of the distance between the lattices $z$ and the lattice spacing $a$, both scaled to wavelength that equals $2\pi$ in the present units. To avoid outright divergences, the $z$ and $a$ coordinates run over a range that does not reach down to $0$ or up to $2\pi$. Bottom: Contour plot of the real part of the same ratio $T(z,a)/\bar T(a)$. The plot only shows the range of values from 0.9 to 1.1, the interval between contours being 0.02. Darker shades indicate decreasing values; the large area at the bottom right is bounded by the contour 1.0.
  }
 \label{FT_3D}
\end{figure}

We illustrate the behavior of the transfer sum in Fig.~\ref{FT_3D}. The range of either argument $z$ or $a$ does not reach down to zero or up to $2\pi$, so there are no outright divergences. However, the onset of the divergence with $z\rightarrow0$ for $z<a$ is clearly visible, and hints of the exceptional behavior for $a=2\pi$ are also discernible in the 3D plot of the absolute value of $T(z,a)/\bar T(a)$ (top panel). For the  most part, though, the value of $T(z,a)/\bar T(a)$ is close to one. This is demonstrated in the contour plot of the real part (bottom panel) that only shows the region of $z$ and $a$ with $0.9<\Re[T(z,a)/\bar T(a)]<1.1$. For $z\gtrsim a$, $\bar T(a)= 2\pi i/a^2$ is a passable approximation for the transfer sum $T(z,a)$. In fact, except for problems associated with Bragg reflections, at large distances $z$ between the lattices, $\bar T(a)$ becomes an increasingly accurate approximation to the transfer sum $T(z,a)$. Moreover, at large distances the radiation from the lattice obviously makes a plane wave. $\bar T(a)d$ is a good approximation for the electric field not only at the sites of a second lattice, but in the far field also in an entire plane parallel to the lattice.

\subsection{Total reflection from a lattice}
\label{TOTREF}

We now combine what we have so far for an investigation of light transmission and reflection for one 2D lattice. Take an incoming field with the amplitude $E_0$ coming perpendicularly to the lattice, then the field $E$ and the dipole moment $d$ at each lattice site satisfy
\beq
E = E_0 + S(a) d,\quad d = \alpha E\,;
\label{TOTALFIELD}
\eeq
see Eq.~\eq{POLARIZABILITY} for the polarizability $\alpha$.
This allows us to solve the dipole moment as a function of the incoming field as
\beq
d =-\frac{3 E_0}{2 [\delta +\threehalves S(a)+i]} = -\frac{3 E_0}{2 (\Delta +i\Gamma)},
\label{dFORM}
\eeq
where
\beq
\Delta = \delta + \threehalves \Re [S(a)], \quad \Gamma = 1 + \threehalves \Im [S(a)]
\eeq
are the effective detuning and transition linewidth. The dipole moment is as for a single atom, except that the cooperative response of the lattice has shifted the resonance  by $-\threehalves\Re[S(a)]$, converting the detuning to the effective detuning  $\Delta$, and the linewidth is changed to $\Gamma$.

Next consider the reflected and transmitted fields far away from the lattice, plane waves such that the long-distance transmission amplitude $\bar T(a)$ applies. We write
\beq
\bar T(a) = i t(a),\quad t(a)=\frac{2\pi}{a^2}\in \mathbb{R}\,. 
\label{TBARA}
\eeq
The total reflected and transmitted amplitudes are then
\bea
E_R &=& d\, \bar T(a) =  \frac{-i\threehalves t(a)}{(\Delta + i \Gamma)}\,E_0\,,\nonumber\\
E_T &=& E_0+d\, \bar T(a)=\frac{\Delta + i [\Gamma - \threehalves t(a) ]}{\Delta + i \Gamma}\,E_0\,.
\label{AMPS}
\eea
Since $\Gamma$ is positive, the transmitted field equals zero if and only if both the effective detuning $\Delta$ equals zero, and the imaginary part of the numerator in $E_T$ is also zero, 
\beq
\Gamma - \threehalves t(a) = 0\,.
\label{CONDITION}
\eeq
This can be cast in the form
\beq
\Im[S(a)] = \frac{2\pi}{a^2}-\frac{2}{3}\,.
\label{IMG}
\eeq
As can be verified by comparing against numerical results, this actually is the dependence of the imaginary part of the self-sum on the lattice spacing in the interval $a\in(0,2\pi)$. Others have shown the same using analytical arguments~\cite{SHA17}.

\begin{figure}[tb]
  \centering
  \includegraphics[width=0.9\columnwidth]{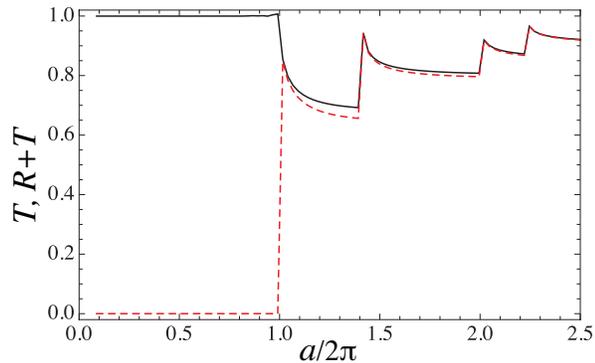} 
  \caption{Power transmission coefficient $T$ (dashed red line) and the sum of reflection and transmission coefficients $R+T$ (solid black line) as a function of lattice spacing $a$. In this figure the effective detuning $\Delta$ is set to zero at each lattice spacing. }
 \label{FRT}
\end{figure}

Thus, while $a\in(0,2 \pi)$ holds true, there is always a detuning $\delta$ such that $\Delta=0$, and no light gets transmitted through the lattice. As may be seen from Fig.~\ref{FS}, there are two values of the lattice constant in the interval $a\in(0,2 \pi)$ with $\Re[S(a)]=0$, and hence for on-resonance excitation with $\delta=0$ also $\Delta=0$ holds true. These are the lattice spacings, approximately $0.2\times 2\pi$ and $0.8\times 2\pi$, for which the lattice does not transmit any resonant light~\cite{Bettles_prl16,SHA17}.

Our development also provides unique insights into the zero transmission, and the full reflection that accompanies it. To begin with, let us define the power transmission and reflection coefficients
\beq
R,T = \left|\frac{E_{R,T}}{E_0} \right|^2\,.
\label{POWS}
\eeq
Naively, $R+T$, the sum of reflection and transmission coefficients, equals one if energy is conserved, as it should be since in steady state all electromagnetic energy coming to each atom also gets radiated away. In Fig.~\ref{FRT} we show the sum $R+T$ (solid black line) and the transmission coefficient $T$ (dashed red line) as a function of the lattice spacing.  The detuning $\delta$ is always chosen so that the effective detuning equals $\Delta=0$. The difference of the two curves, of course, is the reflection coefficient $R$. The sum $R+T$ is not always equal to one, seemingly indicating that energy is not conserved.

The explanation is that $R$ and $T$ only incorporate the energy that propagates as a plane wave in the direction perpendicular to the lattice. However, when nontrivial Bragg scattering is possible for $a\ge2\pi$, there are always Bragg scattered waves present that transport energy to other directions, and these are not included in the sum $R+T$. Compare Figs.~\ref{FS} and~\ref{FRT}: there are discontinuities in the curves at exactly those $a$ for which a new Bragg order emerges.

Now, assume that the dipole moment associated with the lattice shows a generic single-Lorentzian resonance of the form
\beq
d = -\frac{D^2}{\Delta + i \Gamma} E_0,
\eeq
where $D^2$ is a positive constant, $\Delta$ is a measure of detuning from resonance, and $\Gamma$ is the linewidth. Furthermore, assume that in the far field the radiation is out of phase with the dipole moment by $\pi/2$, so the radiated field is of the form
\beq
E_{R} = i t d =  -\frac{i t D^2}{\Delta + i \Gamma} E_0
\eeq
with some positive $t$. On the side of the reflection, this is the total field. On the side of the transmission, one adds the incoming field,
\beq
E_{T} = E_0 + E_{R} = \frac{\Delta + i(\Gamma - t D^2)}{\Delta + i \Gamma} E_0\,.
\eeq
The sum of reflection and transmission coefficients can be manipulated into the form
\beq
R+T =1+ \frac{2tD^2(tD^2 - \Gamma)}{\Delta^2+\Gamma^2}\,.
\eeq
This can be identically equal to one only if $\Gamma=tD^2$. But this is also precisely the condition that there exists a detuning such that the transmission and reflection coefficients are equal to zero and one, respectively.
The other assumptions except a single Lorentzian resonance are generic to atoms interacting with light, and even have obvious counterparts in the general theory of linear response.

In the case of a single Lorentzian resonance energy conservation per se dictates that there must be a detuning for which total reflection prevails for a 2D lattice on normal incidence. Energy conservation works in this way because for plane wave excitation, and in the absence of Bragg scattering, the interference from the lattice atoms forces the radiation from the lattice to propagate either in the direction of  the incoming light, or in the opposite direction. Cooperation between the atoms is not directly needed, although it evidently must be part of a consistent theoretical description. A single Lorentzian resonance is a sufficient condition for total reflection, but it will turn out below, Sec.~\ref{MAGFIELD}, that it is not a necessary condition. 

Related arguments have been made before~\cite{Facchinetti,FacchinettiLong}, but our logic is different. We place the Lorentzian form of the resonance at front and center, which is the cue for our further developments about non-normal incidence in Sec.~\ref{ARBINC}.

\section{Stack of lattices}
\label{STACK}

We next consider the situation when $N$ 2D square lattices with the lattice constant $a$ are stacked so that the $x$ and $y$ coordinates of the atoms are the same for all lattices. The driving light propagates in the perpendicular $z$ direction. The positions of the lattices are denoted by $z_n$. For a regularly spaced stack with the spacing $\Delta z$ we could write $z_n = (n-1)\Delta z$, $n=1,2,\ldots,N$, but almost nothing in our formal development depends on the regular spacing.

We know that a 2D lattice can completely reflect back all of the incoming light, just like a single atom in a 1D channel for light~\cite{Javanainen1999a,RUO16,RUO17}. One would thus think that the existing understanding of atoms in an effectively 1D waveguide could be transferred  to a stack of lattices. This notion is by and large correct. The basic difference is that in a 1D waveguide there is only one independent value of the electric field (for a given polarization mode) at every 1D coordinate, call it $z$. The response of an atom will impose a condition tying the incoming and scattered lights on both sides of the atom, and one can set up a transfer-matrix analysis~\cite{RUO17}. Unfortunately it is not so for a stack of 2D lattices. The electric field falling on the sites of a lattice does not uniquely determine the electric field in the {\em whole\/} plane of the lattice, and therefore neither the field that would excite the next lattice. This will directly and indirectly cause differences between a 1D waveguide and a stack of 2D lattices. Our question is, how much of a difference?

Now, given the electric field $E_m$ at the $m^{\rm th}$ lattice, the lattice atoms develop a dipole moment $d_m$ as before, Eq.~\eq{dFORM}. The lattice then produces a scattered field whose amplitude on the atoms of a lattice at $z_n$ equals
\beq
E_n =e^{i|z_n-z_m|} T(|z_n-z_m|,a)\,  d_m\equiv G(n,m)\, E_m\,.
\eeq
Here we have put back the propagation phases of light that were transformed away in the definition of the transfer sum. The propagator for the light is
\beq
G(n,m) = -\frac{3 T(|z_n-z_m|,a)}{2 [\delta +\threehalves S(a)+i]} e^{i |z_n-z_m|}\,.
\label{PROPAGATOR}
\eeq
Given an incoming plane wave with the amplitude $E_0$, the fields at the sites of the lattices are related by
\beq
E_n = E_0 e^{i z_n} + \sum_{m\ne n} G(n,m) E_m\,.
\label{SSEQ}
\eeq
This is an inhomogeneous set of linear equations for the fields $E_n$, which can be solved. Given the fields, we can compute the dipoles $d_n$, and from them the total transmitted field. Far behind the stack the long-distance form of the transfer sum $\bar T(a)$ is operative, and at large $z$ we have the transmitted field
\beq
E(z) = E_0 e^{iz} + \sum_n G_0(z,m) E_m,
\label{FARFIELD}
\eeq
with
\beq
G_0(z,m) =  -\frac{3 \bar T(a)}{2 [\delta +\threehalves S(a)+i]} e^{i|z-z_m|}\,.
\eeq
Transmission and reflection coefficients of the stack of lattices may be computed from the far field~\eq{FARFIELD}.

Suppose the lattice spacing satisfies $a\in(0,2\pi)$, so that~\eq{CONDITION} holds true. We may write Eq.~\eq{PROPAGATOR} in the form
\beq
G(n,m) = -i\frac{\Gamma \zeta(n,m)}{\Delta+i \Gamma} e^{i\Phi(n,m)} e^{i|z_n-z_m|}.
\label{LATPROP}
\eeq
Here $\zeta>0$ and $\Phi\in \mathbb{R}$ are the modulus and the phase of the  complex ratio that characterizes the difference between the transfer sum and its long-distance form,
\beq
 \zeta(n,m) e^{i \Phi(n,m)}= \frac{T(|z_n-z_m|,a)}{\bar T(a)} .
\eeq

The counterpart of the propagator~\eq{LATPROP} for atoms in a 1D waveguide~\cite{RUO16,RUO17} is, in the present units,
 \beq
 G_{1D}(n,m) = - \frac{i\zeta}{\delta + i}\,e^{i|z_n-z_m|}\,.
 \eeq
 Here $\zeta$ is the fraction of the light energy emitted by an atom that ends up back in the waveguide, as opposed to leaking to free space. The cooperative resonance frequency and linewidth of the lattice, instead of the resonance quantities of a single atom, are the obvious differences between a lattice and an atom in a 1D waveguide. Moreover, in a 1D waveguide there would be no adjustments  $\Phi$ to the propagation phases, and, in lieu of the factors $\zeta(n,m)$ that depend on the positions of both lattices, a (homogeneous) 1D waveguide would have a constant $\zeta$. For an ideal waveguide, $\zeta=1$.

Since $T(|z_n-z_m|,a)\rightarrow \bar T(a)$ when either $a\rightarrow0$ or $|z_n-z_m|\rightarrow \infty$,  both of these limits produces a simulacrum of a perfect 1D waveguide. As far as light transmission through a 1D waveguide goes, the distances between the successive atoms only matters modulo half of a wavelength, $\pi$ in our present units~\cite{RUO16,RUO17}. Thus, by placing the lattice planes far enough apart, it is in principle possible to emulate the physics of an arbitrary string of atoms in a perfect 1D waveguide to an arbitrary precision using a stack of 2D lattices.

\begin{figure}[tb]
  \centering
  \includegraphics[width=0.9\columnwidth]{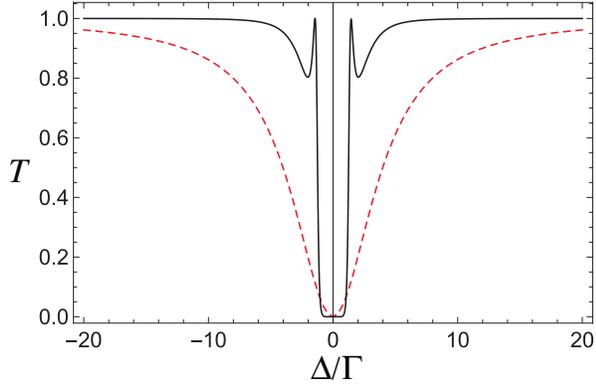} 
  \caption{Power transmission $T$ as a function of the scaled detunings $\Delta/\Gamma$ for a stack of $N=4$ lattice planes with both the lattice constant $a$ and the stack spacing $\Delta z$ being $\pi/2$ (solid black line), and $\pi$ (dashed red line). }
 \label{TRMS}
\end{figure}

The final question is about the quality of the emulation. In Fig.~\ref{TRMS} we plot the transmission as a function of the scaled detuning $\Delta/\Gamma$ for a stack of $N=4$ lattices. Figures 5 and~6 of Ref.~\cite{RUO17} shows analogous graphs for a 1D waveguide, though none of them for the exact same parameter values.

The scaling $\Delta/\Gamma$ removes the cooperative resonance shift and broadening of each 2D lattice from consideration. The dashed red line is for the geometric constants $a=\Delta z=\pi$, half of a wavelength, and shows the superradiant resonance with the width $4\, \Gamma$.  The solid line is for the constants  $a=\Delta z=\pi/2$. It displays a sharp stop band with two narrow transmission features due to Fano resonances in the cooperative response of the {\em lattices\/} in the stack. As far as we can tell numerically, these resonances still have exactly unit transmission, but the positions are slightly shifted from what one would expects on the basis of atoms in an ideal 1D waveguide: $\Delta/\Gamma \simeq \pm 1.45$ versus $\delta=\pm\sqrt{2}\simeq\pm1.41$. To an unaided eye, the graphs in Fig.~\ref{TRMS} would be indistinguishable from the corresponding graphs for an ideal 1D waveguide.

\begin{figure}[tb]
  \centering
  \includegraphics[width=0.9\columnwidth]{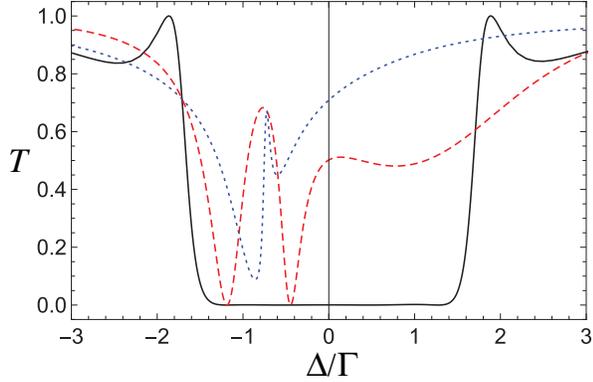} 
  \caption{Power transmission $T$ as a function of the scaled detunings $\Delta/\Gamma$ for a stack of $N=4$ lattice planes at the fixed stack spacing~$\Delta z = \pi/2$, for different lattice constants: $a = \pi$ (solid black line), $a=1.8\pi$ (dashed red line) and $a=2.2\pi$ (dotted blue line). }
 \label{VZTRMS}
\end{figure}

Close to $a=2\pi$, starting at about 15\% below a wavelength, and on both sides of $a=2\pi$, we see substantial deviations from the picture of an ideal 1D waveguide, even after scaling away the cooperative effects of each individual 2D lattice by plotting the results as a function of $\Delta/\Gamma$. An example is provided in Fig.~\ref{VZTRMS}. This shows the transmission for a stack of four lattices at the fixed spacing $\Delta z= \pi/2$, for different lattice constants. At $a=\pi$ , solid black line, the result is similar to the result shown in Fig.~\ref{TRMS} for $a=\Delta z= \pi/2$, but the other two graphs for $a = 1.8\, \pi$  (dashed red line) and $a=2.2\,\pi$ (dotted blue line) deviate from the corresponding 1D fare substantially.

We seek further illumination from time evolution. Let us denote the total field strength at lattice $n$ by $E_n$, then the linear dipole at site $n$ satisfies the time evolution equation
\beq
\dot d_n = (i\delta -1) d_n + \threehalves i E_n.
\eeq
The steady-state solution gives $d_n = \alpha E_n$ with the polarizability~\eq{POLARIZABILITY}, as it should. We also state the assumption that the propagation time of light across the sample is much smaller than the response time of the atoms, something like the inverse of the linewidth.  Then the field $E_n$ satisfies
\beq
E_n = E_0 e^{i z_n} + S(a) d_n + \sum_{m\ne n} T(|z_n-z_m|,a) e^{i|z_n-z_m|} d_m
\eeq
as an instantaneous equation.
This gives a coupled equation for the dipoles
\bea
\dot d_n &=& (i\Delta -\Gamma) d_n + \threehalves i \sum_{m\ne n} T(|z_n-z_m|,a)e^{i|z_n-z_m|} d_m 
\nonumber\\ &+&\threehalves i E_0 e^{i z_n}\,.
\label{TIME}
\eea

\begin{figure}[tb]
  \centering
  \includegraphics[width=0.96\columnwidth]{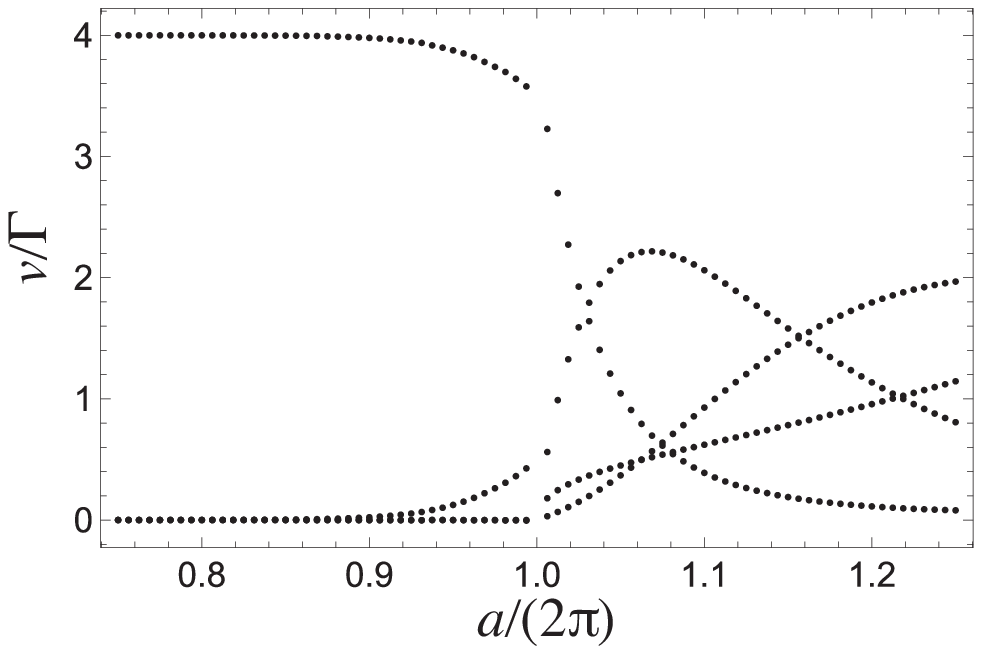} \\
  \hspace{8pt} \includegraphics[width=0.985\columnwidth]{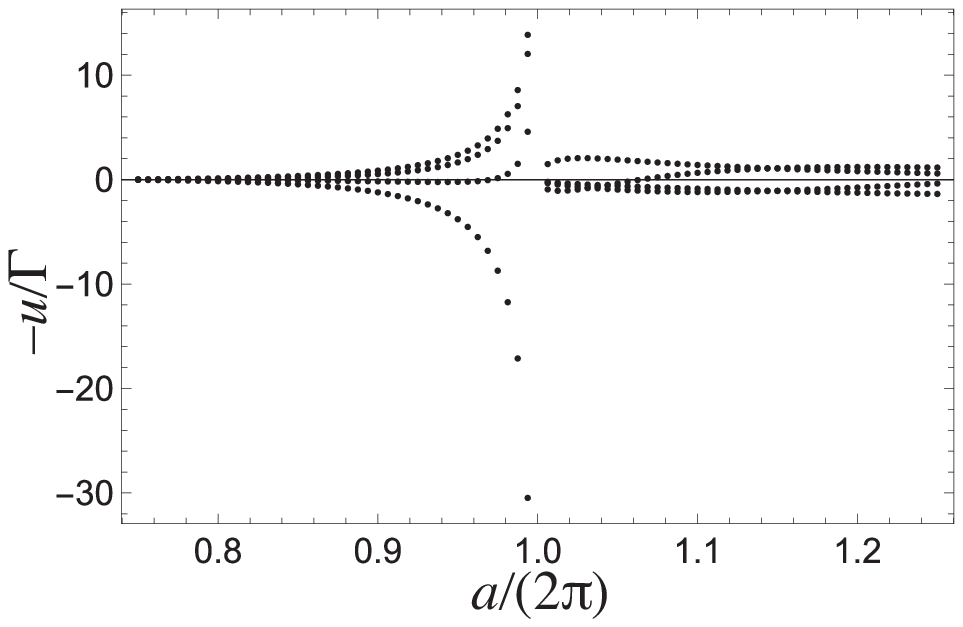} 
  \caption{Damping rates (top) and resonance shifts (bottom) of the collective eigenmodes of a stack of four lattices as a function of the lattice constant $a$. The spacing of the lattices is a constant, $\Delta z=\pi$. }
 \label{EVFIG}
\end{figure}

The inhomogeneous version of Eq.~\eq{TIME} with $E_0=0$ gives the cooperative radiative modes of the stack of lattices. Specifically, with the ansatz $d_m(t)= \tilde d_m  e^{-i \omega t} \equiv e^{(i u - v)t}$, we have the eigenvalue equation
\beq
-i \omega \tilde d_n = (i\Delta -\Gamma) \tilde d_n + \threehalves i \sum_{n\ne m} T(|z_n-z_m|,a)e^{i|z_n-z_m|} \tilde d_m \,.
\eeq
The effective detuning and linewidth for each mode, $u$ and $v$, incorporate the cooperative effects from the radiative coupling between the lattices. As usual, the corresponding eigenvectors made of $\tilde d_n$ determine the coupling of the eigenmodes to the driving plane wave.

Once more, in the case $a\in(0,2\pi)$ and under the generally quite good an approximation $T(|z_n-z_m|,z)= T(a)$, the time dependent collective modes and collective behavior of the stack of lattices would be the same as the behavior of the atoms in a lossless 1D waveguide, save for the lattice shift and line broadening in the detuning $\Delta$ and linewidth $\Gamma$. 

However, $T(z,a)\simeq \bar T(a)$ breaks down around $a\simeq 2\pi$. What happens upon approach, and crossing, the limit $a=2\pi$ is shown in Figs.~\ref{EVFIG}. This is for a stack of $N=4$ lattices, for the constant spacing between the lattices $\Delta z=\pi$. The corresponding ideal 1D waveguide would have four eigenmodes with no resonance shifts, $u=0$, whereas one eigenmode would have a superradiant decay rate $v=4$ and the other three would have the extreme subradiant decay rate $v=0$~\cite{RUO17}. The broad resonance in Fig.~\ref{TRMS} (dashed red line) originates from the coupling of the driving light to the superradiant mode.

Figure~\ref{EVFIG} shows the decay rates (top) and resonance shifts (bottom) of the eigenmodes as a function of the lattice constant $a$. The detuning is chosen so that $\Delta=0$. The results are presented in units $\Gamma$, and the effects of the 2D lattice per se  have again been renormalized away. The way we have defined the detuning, the cooperative resonance shift actually equals $-u$, as given in the figure. Below but up to quite close to  $a=2\pi$, the analog of the 1D waveguide prevails. The behavior for $a>2\pi$ is different, but the damping rates (in the units $\Gamma$, which is singular in itself) appear to interpolate continuous across the Bragg reflection singularity at $a=2\pi$. 

In summary, a stack of 2D lattices generally speaking  behaves closely like a 1D ideal waveguide with embedded atoms, although these quasiatoms have the resonance frequency and the radiative linewidth of the 2D lattice. The notable exceptions we have demonstrated occur at lattice spacings such that 2D Bragg scattering is, or is about to become, a consideration.

\section{Arbitrary angle of incidence}
\label{ARBINC}
Up to now, the light has always propagated in the direction perpendicular to the lattice. We now take up the case of an arbitrary direction of the incoming light.

\subsection{Formulation}
Henceforth we express the direction of the incoming plane wave using spherical polar coordinates. With our choice of the units the length of the wave vector of the incoming light equals one, so we write
\beq
{\bf k} ={\bf k}(\theta,\phi)= \sin\theta \cos\phi\, \hat{\bf e}_x +  \sin\theta \sin\phi\, \hat{\bf e}_y +  \cos\theta\, \hat{\bf e}_z\,,
\label{KVECTOR}
\eeq
$\theta=0$ means perpendicular incidence, and we assume that the incoming light propagates from ``left'' to ``right'' so that $0\le\theta<\pi/2$. The complex polarization vector of the incoming beam is denoted by $\hat{\bf e}$. For a valid incoming plane wave it must be that ${\bf k}\cdot\hat{\bf e}=0$. Here the dot $\cdot$ stands for the inner product as for real vectors.

The obvious effect of the tilt is to cause a rolling phase  for both the electric field and the dipoles according to the projection of the wave vector on the plane of the atoms, $\bk_\parallel =  \sin\theta \cos\phi\, \hat{\bf e}_x +  \sin\theta \sin\phi\, \hat{\bf e}_y$, but a single vectorial amplitude still characterizes both the dipole moment and the electric field at the positions of the atoms. We write the dipolar field, excluding the self-field, at the center point of the lattice in terms of a self-sum tensor,
\beq
\bE(0)= {\sf S}\bd\,,\qquad {\sf S}(a,\bk_\parallel) = \sum_{{\bf n}\ne 0} e^{i\bk_\parallel\cdot{\bf R_{\bf n}}} {\sf G} (-{\bf R}_{\bf n})\,.
\eeq
The matrix $\sf S$ is symmetric. Moreover, symmetry considerations suggest, and explicit calculations confirm, that the $z$ components of the light field and of the dipoles do not couple to the $x$ and $y$ components, giving ${\sf S}_{xz}={\sf S}_{yz}=0$.

Likewise, the field transferred to the center site of a lattice at the distance $z$ away is characterized by what we call (somewhat confusingly, given the standard terminology) the transfer matrix $\sf T$,
\beq
{\sf T} (z,a,\bk_\parallel )= e^{-ik_\perp|z|}\sum_{\bf n} e^{i\bk_\parallel\cdot{\bf R_{\bf n}}} {\sf G} (z\hat{\bf e}_z-{\bf R}_{\bf n})\,.
\label{TRFMATRIX}
\eeq
The purpose of the exponential prefactor is to cancel the free-propagation phase over the distance between the planes; $k_\perp =  \sqrt{1-\bk_\parallel^2}=\cos\theta$. The matrix $\sf T$ is symmetric as well. The behavior of the tensor ${\sf T}$ under the reflection $z\rightarrow -z$ warrants caution: the $z$ component of the electric field radiated by the components of the dipoles that lie in the plane of the atoms have opposite signs on the opposite sides of the plane, but the field from the $z$ components of the dipoles is the same at both $z$ and $-z$. This symmetry implies that ${\sf T}_{xz}(z) = -{\sf T}_{xz}(-z)$ and  ${\sf T}_{yz}(z) = -{\sf T}_{yz}(-z)$, but these and their matrix-symmetry counterparts are the only elements of ${\sf T}$ that change upon the reflection $z\rightarrow-z$.

It turns out that the transfer matrix ${\sf T}$ also has a large-distance/dense-lattice limit analogously to the transfer sum $T$. By comparing with numerical computations we have found that for positive $z$ the limit is
\beq
\bar{\sf T} = \frac{2\pi i}{a^2}\,{\sf R}(-\phi) {\sf M}(\theta)  {\sf R}(\phi),
\eeq
with
\bea
{\sf M}(\theta) &=& 
\left[
\begin{array}{ccc}
\cos\theta & 0  & -\sin\theta  \\
  0&  \frac{\displaystyle1}{\displaystyle\cos\theta} &  0 \\
 -\sin\theta & 0  &    \frac{\displaystyle\sin^2\!\theta}{\displaystyle\cos\theta}
\end{array}
\right],\quad
\label{MATRICES}
\\
{\sf R}(\phi) &=& 
\left[
\begin{array}{ccc}
 \cos\phi &\sin\phi   &  0 \\
 -\sin\phi & \cos\phi  & 0  \\
0  &  0 &   1
\end{array}
\right];
\label{ROTMAT}
\eea
see Appendix~\ref{NUMDET-T}.

Particularly interesting from our viewpoint is that the large-distance limit of the transfer matrix may also be written as 
\beq
\bar{\sf T} = \frac{2\pi i}{a^2\cos\theta}\,{\sf P}_\perp({\bf k}),
\label{TBAR}
\eeq
where ${\sf P}_\perp({\bf k})$ stands for the projector to the subspace orthogonal to the vector ${\bf k}$. This is how the formalism enforces the condition that the reradiated field is transverse. When  the lattice is viewed at the angle $\theta$ away from perpendicular, the apparent area density of the dipoles is $\bd/(a^2\cos\theta)$. This neatly explains the  $\cos\theta$ in the denominator of~\eq{TBAR}.

Bragg reflections are a possibility for oblique incidence, even more so than for normal incidence. This is discussed in detail in Appendix~\ref{BRAGREF}. However,  unless Bragg reflections are specifically mentioned, we continue to assume that they are absent.

\subsection{Single plane of atoms}

Special cases proliferate when we allow non-perpendicular incidence. We avoid detailed descriptions of quantitative results, and concentrate on prominent qualitative features of reflection and transmission from a single plane of atoms. Our observations are extracted from, or supported by, numerical calculations. Although we do not elaborate on this, the matrix $\sf S$ inherits symmetries from the underlying square lattice that help to discover the results. It is not an accident that we occasionally call the matrices $\sf S$   and $\sf T$ ``tensors.''

We have the incoming field $\bE_0$ and the total field $\bE$ at the center site related by
\beq
\bE = \bE_0 + {\sf S}{\sf A}\bE,
\eeq
where $\sf A$ is the polarizability tensor. The response of an atom with the $J=0\rightarrow J'=1$ transition in the absence of an external magnetic field is isotropic, and $\sf A$ simply equals polarizability times the unit matrix. However, we retain the option of anisotropic atomic response in our formalism for applications below, in Sec.~\ref{MAGFIELD}. It is  a simple matter to solve the total field and the atomic response,
\beq
\bE = (1-{\sf{SA}})^{-1}\bE_0,\quad \bd={\sf A} (1-{\sf{SA}})^{-1}\bE_0.
\label{BEQEQ}
\eeq
We write the reflected and transmitted fields on the left ($z<0$) and on the right ($z>0$) in the form
\bea
\bE_R &=& e^{-ik_\perp z } {\sf T}(-z){\sf A} (1-{\sf{SA}})^{-1}\bE_0, \\
\bE_T &=& e^{ik_\perp z}[1+ {\sf T}(z){\sf A} (1-{\sf{SA}})^{-1}]\bE_0\,.
\label{ENDEQ}
\eea 

In the rest of this section we always use the large-$z$ limit $\bar{\sf T}$ for the transfer matrix. The reflected field propagates in the direction
\beq
{\bf  k}_R = \sin\theta \cos\phi\, \hat{\bf e}_x +  \sin\theta \sin\phi\, \hat{\bf e}_y -  \cos\theta\, \hat{\bf e}_z\,.
\eeq
Correspondingly, the $z\rightarrow-\infty$ limit of the transfer matrix $\bar{\sf T}_R$ projects to the subspace orthogonal to ${\bf  k}_R$.

So far we take the polarizability tensor $\sf A$ to be diagonal, corresponding to the usual isotropic polarizability $\alpha$ as in~\eq{POLARIZABILITY}.  The induced dipole is then
\beq
\bd = -\threehalves(\delta + i + \threehalves {\sf S})^{-1} \bE_0.
\eeq
The matrix $\sf S$ has three eigenvalues, and (without proof) we expect it to have three linearly independent eigenvectors. The component of the electric field along the eigenvector corresponding to the eigenvalue $s_i$ drives a resonance with the effective detuning and linewidth
\beq
\Delta_i = \delta + \threehalves\Re[s_i],\quad\Gamma_i = 1+\threehalves\Im[s_i]\,.
\eeq

Now, take any incoming beam of light characterized by the polarization vector $\hat{\bf e}$. The polarization vector may be decomposed into a linear superposition of the eigenvectors of $\sf S$, and there are in general three separate resonances with separate resonance frequencies and linewidths that show up in the response; one for each eigenvector. The matrix $\sf S$ has no elements that couple the $z$ polarization to any polarization in the plane, so at this stage the $z$ components of the dipoles and of the electric field, if any, get processed separately. In the subsequent reradiation stage governed by the transfer matrices $\bar{\sf T}$ and $\bar{\sf T}_R$ the eigenvectors are projected to the two-dimensional subspaces of transverse polarizations allowed for the exiting plane waves, but in the most general case the outgoing light still carries along behavior from all three resonances.

For normal incidence, $\theta=0$, there is no difference between the $x$ and $y$ directions, and the incoming field must be polarized in the plane of the atoms. Therefore ${\sf S}_{xx}={\sf S}_{yy}$ and ${\sf S}_{xy}~=~0$ must hold true, and the dipolar resonance in the $z$ direction cannot be excited. Since the $x-y$ part of the matrix ${\sf S}$ is diagonal, {\em all\/} polarizations in the $x-y$ plane are eigenvectors with the same eigenvalue. We are back to the single-resonance case we have  discussed already.

For non-normal incidence there in general is a $z$ component in the polarization of the incoming light, which is subject to its own separate resonance. For a general propagation direction there are also two elliptically polarized eigenmodes of $\sf S$ in the $xy$ plane, which are linearly independent though not necessarily orthogonal  in the sense of the proper inner product for complex vectors. We expect two resonances in the $x-y$ plane as a result. The situation simplifies if the projection of the propagation vector $\bk_\parallel$ to the lattice plane is along a direction of high symmetry. If $\bk_\parallel$ lies in a direction of a lattice axis ($\phi=0,\,\pi/2,\,\ldots$), there are two resonances linearly polarized along the axes. If $\bk_\parallel$ bisects an angle between the lattice axes  ($\phi=\pi/4,\,3\pi/4,\,\ldots$), there are likewise two eigenvectors of $\sf S$ in the plane polarized parallel and perpendicular to $\bar\bk_\parallel$.

Taking into account the transversality of light, we have two guaranteed ways to produce just a single resonance in a square lattice of atoms with oblique incidence: Choose the propagation vector of light in such a way that its component in the lattice plane either lies in the direction of a lattice axis or bisects the axes, and choose the linear polarization of the light in such a way that it lies in the lattice plane. This simplification is based on the symmetries. There may be other symmetries or just plain accidental special cases of a single resonance, but we have not run into any.

At this point the question of whether it is possible to attain zero transmission or perfect reflection for non-normal incidence practically asks itself. First, even in the normal-incidence case perfect reflection can only happen on exact lattice-shifted resonance. We therefore hypothesize that perfect reflection from a single lattice is not possible if more than one resonance gets excited, as the frequency of the light could not be the right one for all resonances at the same time. Our experience supports this hypothesis --- the difference from perfect reflection may be small, but we have always been able to detect it. Moreover, we have studied  numerically transmission of the eigenmodes of $\sf S$ regardless of whether they are proportional to legal polarizations. We surmise that complete reflection is only possible if the eigenmode, in fact, is a valid transverse polarization for a plane wave, and gets projected untruncated by the transfer matrices $\sf T$ and ${\sf T}_R$. These considerations together suggest that the special cases of the preceding paragraph are precisely the cases when one can reach complete reflection. A magnetic field that causes additional mixing of in-plane and out-of-plane polarizations  (Sec.~\ref{MAGFIELD}) changes things, but otherwise our numerical experiments support this hypothesis.

\begin{figure}[tb]
 \centering
 \includegraphics[width=0.9\columnwidth]{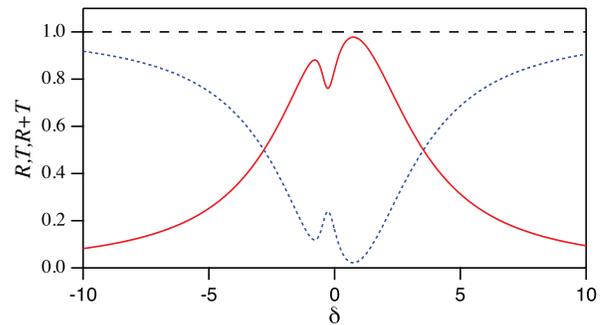}
 \caption{Reflection coefficient $R$ (solid red line), transmission coefficient $T$ (dotted blue line), and the sum of the two $R+T$ (dashed black line) as a function of detuning for non-perpendicular incidence. The parameters are $a=\pi$, $\theta=0.4\,\pi$, and $\phi=0.125\,\pi$, and the light is polarized in the plane of the atoms.}
 \label{TENEX}
\end{figure}

These concepts are illustrated in Fig.~\ref{TENEX} showing the transmission and reflection coefficients along with the sum ($R+T=1$) of the two for non-perpendicular incidence. The lattice spacing here is $a=\pi$, half of the wavelength, the light comes in at an oblique angle of $0.4\,\pi$ ($\theta=0$ means perpendicular incidence and $\theta=0.5\,\pi$ would be light coming in along the lattice plane), and the projection of the wave vector onto the lattice plane makes the angle $\phi=0.125\,\pi$ with the $x$ axis, i.e., bisects the angle between the $x$ axis and the direction that, in turn, bisects the angle between the $x$ and $y$ axes. The polarization of the light is in the plane of the atoms. The two resonances with the corresponding positions and widths $-0.325+0.389\, i$ and $0.399+3.00\,i$ are obvious in the figure, as is the fact that the reflection coefficient never reaches one.

Our computations show that, as long as there are no Bragg reflections, the sum of transmission and reflection coefficients again equals one. It should be the case by virtue of energy conservation, but we find it remarkable how much of the mathematics of the tensors $\sf S$ and $\sf T$ must align just right to produce this result. If one grants energy conservation, the perfect reflection in the case of the polarization of the light that is a joint eigenvector of $\sf S$ and $\sf T$ is presumably a simple extension of the argument in the 1D case for a single Lorentzian resonance. One could then even find the corresponding damping rates $\Gamma_i$ as a function of the lattice spacing $a$ analytically, analogously to the 1D case.

\section{Canceling light shifts in a lattice}
\label{CANCELLIGHTSHIFTS}
It is a commonly held notion in high-precision spectroscopy that binding the atoms to a lattice will eliminate collisions, and thereby the associated collision shifts. This idea need not be an unqualified success~\cite{CHA04}. We have already demonstrated cooperative shifts of the resonance originating from the dipole-dipole interactions of the atoms in a 2D lattice in Fig.~\ref{FS}. The same dipole-dipole interactions are largely responsible for collision shifts~\cite{LEW80}. Moreover, the analysis of  regularly spaced atoms in a 1D waveguide shows possibly very large shifts of the resonances when the spacing between the atoms is close to half of the wavelength~\cite{RUO17}, which would be the lattice spacing if one could use light close to resonance to make an optical lattice to hold the atoms. This is because the atoms effectively delimit cavities, which pull the resonance. At present we view  a 3D lattice as a 1D stack of 2D lattices, and argue that these 1D and 2D shifts may cancel if the geometry of the experiment is chosen properly.

We first revisit the relevant material in 1D~\cite{RUO17}, so far in fully dimensional quantities. The lattice has the spacing $d$. In addition there is the retention ratio $\zeta$, as also described in Sec.~\ref{STACK}. In the limit of a large number $N$ of atoms and a small retention ratio $\zeta$ such that $N\zeta\lesssim 1$ (whereupon the optical thickness of the sample is at most comparable to unity), the resonance of a 1D waveguide experiences a cavity shift
\beq
s_C = \half \zeta\gamma \cot 2kd\,.
\label{1DSHIFT}
\eeq
It may be very large for an inopportune lattice spacing.

Next return to 2D and 3D lattices. If every site is occupied, on resonance even a very small sample is opaque, and in addition to the cooperative line shifts there could be large line broadenings as a result of the optical thickness. Having only a small fraction of the lattice sites occupied at random might  be an advantage. 

Unfortunately, exact modeling of the random atomic positions most likely requires direct numerical simulations. We attempt to get away with an ad-hoc model. Suppose the filling factor of the lattice is $\zeta$. If we, on the other hand, multiply the polarizability by an adjustable constant $\zeta$, $0<\zeta\le1$, the effect is to  reduce the induced dipole moments by the factor $\zeta$, as if the area density and volume density of the atoms were both reduced by the factor $\zeta$. We equate these two roles of the constant $\zeta$, which is a mean-field type approximation.

 As the final ingredient of the model, we use the large-distance limit of the transfer sum $T$ or the transfer matrix $\sf T$. This effectively means that, in transmission of light, a lattice behaves as if the dipole moment were distributed  continuously over the lattice plane. This is also a mean-field approximation.

Now consider a stack of 2D lattices with the lattice constant $a$ and lattice spacing $d$ a normal incidence. When the polarizability is reduced by the factor $\zeta$, this is just as in a 1D waveguide holding atoms with the fraction $\zeta$ of the radiated energy retained; compare with Sec.~\ref{STACK}. Reverting to the units of the present paper, such atoms in a 2D lattice produce an atom-like response with the linewidth $\Gamma = 1 + \threehalves\zeta\, \Im [S(a)]$ and lattice-induced resonance shift $s_L= -\threehalves\zeta\, \Re [S(a)]$. Let us assume that the filling factor is small enough that $\zeta \, \Im [S(a)]$ is small compared to unity, so the linewidth stays approximately unchanged. By comparing with Eq.~\eq{1DSHIFT}, we have the total shift at normal incidence
\beq
s_\perp = \zeta\left(-\threehalves\, \Re [S(a)] + \half\cot 2d\right).
\eeq

Next move on to non-normal incidence, applying the development in Sec.~\ref{ARBINC}. If the purpose of putting the atoms in a lattice is to improve precision of spectroscopy, one would not want to introduce two or even three separate resonances with their own shifts. We assume that the non-normal incidence is in a configuration for which only one resonance with the complex eigenvalue $s$ gets excited. Besides, incidence at the angle $\theta$ reduces the effective wave number for propagation from lattice plane to lattice plane by the factor $\cos\theta$. We thus have an expression for the resonance shift
\beq
\frac{\delta_T}{\zeta} = -\threehalves\, \Re[s] + \half\cot(2 d \cos\theta).
\eeq

\begin{figure}[tb]
 \centering
 \includegraphics[width=0.9\columnwidth]{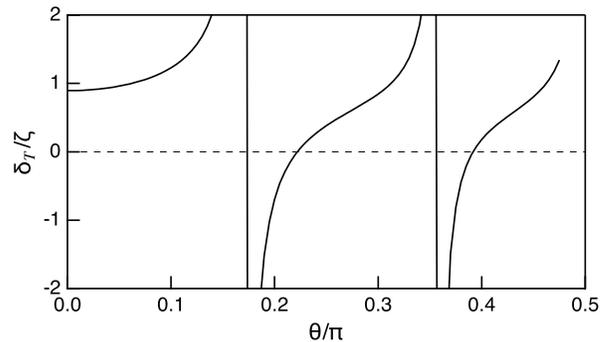}
 \caption{Total density shift of resonance in a 3D lattice as a function of tilt angle in a case modeling a cubic lattice for strontium atoms made with magic-wavelength light; see text for details.}
 \label{TOTSH}
\end{figure}

For a numerical illustration we consider a simple cubic lattice with $d=a$. The lattice spacing equal to half of the magic wavelength $813\,{\rm nm}$ in strontium with the transition wavelength $698\,{\rm nm}$, which gives $a=3.66$. We direct the light so that the projection of the propagation direction bisects the angle between lattice axes, $\phi=\frac{1}{4}\,\pi$. The total resonance shift $\delta_T$ as a function of the angle of incidence $\theta$ is plotted in Fig.~\ref{EVFIG}. There are two angles where the shift gets very large owing to the cotangent behavior of a 1D lattice, but also two angles at which the shift cancels. In the other case of a single resonance with $\bk_\parallel$ along a lattice axis, an angle of incidence with no density shift exists only after (2D) Bragg scattering would have set in already. 

Our argument could be questioned on various grounds. In particular,  we have earlier produced an example~\cite{JavanainenMFT} in which mean-field theory for atom-light interaction fails by a factor of ten.  We break up this issue in two parts: Can a partly filled lattice plane be thought of as kind of a  superatom with mean-field parameters? Is mean-field theory sufficient to describe propagation of light between the superatoms? 

Examples we have worked out numerically in Appendix~C suggests that the mean-field approximation for a 2D lattice is semiquantitatively valid. Two factors may help here. First, in a typical application the lattice spacing is about half of a wavelength, and the density expressed as a dimensionless number is at most $\sim(1/2)^3\sim0.1$. One can detect deviations from mean-field theory even at such ``low'' densities~~\cite{Javanainen2014a,JAV17,JavanainenMFT}, but no qualitative failure. Second, the lattice sets a minimum distance between the atoms and thereby limits the contributions from the $1/r^3$ and $1/r^2$ parts of the dipole-dipole interaction. 

In a 3D cubic lattice the self-sum and transfer sum are very similar in structure. Encouraged by the success of the mean-field approximation within a plane, we therefore surmise that the mean-field theory for light propagation between the planes also remains valid even if there are empty sites. Overall, the errors coming with the mean-field approach should not be large enough to invalidate our conclusions.

While putting atoms in a lattice will not automatically get rid of dipole-dipole interactions and the ensuing analog of collision shifts, a judicious choice of the experimental parameters, including the orientation of the lattice, could eliminate density dependent resonance shifts when the atoms are held in an optical lattice.

\section{Magnetic field}\label{MAGFIELD}
In the present section we consider the effects of the magnetic field on the response of the lattice. The studies in this direction were initiated in Refs.~\cite{Facchinetti,FacchinettiLong}, which propose to use 2D atom lattices for magnetometry. Many of the results we bring up were already introduced in these papers, although the reasons for the results in our treatment are occasionally different in subtle ways.

The technical item of the present section is to give up on isotropic polarizability. The polarizability tensor $\sf A$ may be obtained from quantum mechanics for an arbitrary atomic level scheme and direction of the magnetic field, a computation that is implemented numerically as part of the ``Software Atom''~\cite{JAV17SA}. For a $J=0\rightarrow J'=1$ transition it  may even be obtained  from a classical analysis of the motion of a charge bounded to a harmonic oscillator potential, and acted on by both a sinusoidally oscillating electric field and a static magnetic field. We outline this argument in Appendix~\ref{POLAPP}. The numerical parameter that encapsulates the strength of the magnetic field,  $\omega_B$, equals the Zeeman splitting in the excited level in units of the linewidth. It is defined here based on the frequency difference between the states $m'=1$ and $m'=0$, and is negative if the Lande factor of the level $J'=1$ is positive.

Now that the polarizability is a nontrivial $3\times3$ matrix, we need to use the full formalism for the reflected and transmitted fields as in Eqs.~\eq{BEQEQ}-\eq{ENDEQ}, and in the most general case we may have to consider an arbitrary direction and polarization of the incident field as well. We have found in numerical examples  with full sets of basically random parameters that, in the absence of Bragg scattering, energy is again conserved, $R+T=1$.

We discuss as an example the special case when the light is incident perpendicularly to the atomic plane and is linearly polarized in the $x$ direction, while the magnetic field is in the $y$ direction. The motion of the charges induced by the magnetic field is in the $z$ direction, perpendicular to the atomic plane and in the direction of propagation of light. The dipole moment will have a component in the $z$ direction as well.

Facchinetti et al.~\cite{Facchinetti,FacchinettiLong} have studied this situation using the eigenvalues and eigenvectors of the joint atom-field system along similar lines as in our discussion around Eq.~\eq{TIME}. They note the existence of a subradiant mode in which the atoms store the electromagnetic field for extended periods of time. The proximate reason is that the atoms get polarized in the direction of propagation of the light, whereupon they do not radiate. The authors also introduce a two-mode model that, with proper extrapolation to the limit of infinite lattice, is mathematically equivalent to our method.

\begin{figure}
\vspace{15pt}
\includegraphics[width=0.9\columnwidth]{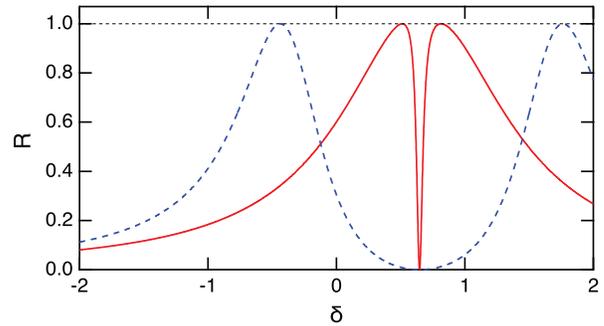}
 \caption{Reflection coefficient $R$ for perpendicular incidence as a function of the detuning $\delta$ for two values of the Zeeman splitting $\omega_B = -0.15$ (red solid line) and $\omega_B=-1.1$ (blue dashed line). The magnetic field is in the plane of the lattice of the atoms, and is also perpendicular to the polarization of the incident field. In this figure the lattice spacing is $a = 0.55\times 2\pi$.}
\label{FACCFIG}
\vspace{-10pt}
\end{figure}

As an example, consider Fig.~\ref{FACCFIG}. We set the lattice spacing $a=1.1\,\pi$ and the Zeeman splitting $\omega_B = -0.15$ (solid red line) or $\omega_B = -1.1$ (blue dashed  line), and compute the reflection coefficient numerically. This figure is our counterpart of the right panel of Fig.~4 in \cite{Facchinetti}. In particular, the dip in both curves indicates excitation of the subradiant cooperative mode. Unlike in the figures in Refs.~\cite{Facchinetti,FacchinettiLong} showing results for finite-size lattices, in our results for an infinite lattice the dips in both curves reach right down to zero. This indicates no reflection, and perfect transmission. Also, as illustrated with the help of the dotted black line at $R=1$, the reflection coefficient always (for $\omega_B\ne0$) touches unity at two detunings; one on each side of the detuning giving the complete transmission.

We may easily gain semi-analytical insights into the behaviors of reflection and transmission. To begin with,  neither the polarizability tensor $\sf A$ nor the self-sum tensor $\sf S$ couple the $y$ components of vectorial quantities to the $x$ and $z$ components, so we only need a two-dimensional description. The required tensors are
\begin{widetext}
\beq
{\sf A} = \left[
\begin{array}{cc}
{\sf A}_{xx}&{\sf A}_{xz}\\
{\sf A}_{zx}&{\sf A}_{zz}
\end{array}
\right]
= \frac{3}{4}\left[
\begin{array}{cc}\displaystyle
 -\frac{\hbox{\small 1}}{\delta +\omega _B+i}-\frac{\hbox{\small 1}}{\delta -\omega
   _B+i} &\displaystyle \frac{ i}{\delta -\omega _B+i}-\frac{ i}{
  \delta +\omega _B+i} \\
\displaystyle \frac{ i}{ \delta +\omega _B+i}-\frac{i}{\delta -\omega
   _B+i} &\displaystyle -\frac{\hbox{\small 1}}{\delta +\omega _B+i}-\frac{\hbox{\small 1}}{\delta
   -\omega _B+i} \\
\end{array}
\right], \quad
{\sf S} =  \left[
\begin{array}{cc}
{\sf S}_{xx}&0\\
0&{\sf S}_{zz}
\end{array}
\right],
\eeq
where the elements of the self-sum tensor $\sf S$ need to be computed numerically, and depend only on the lattice spacing $a$. We have the complete solution, electric field amplitudes and dipole moments, from Eqs.~\eq{BEQEQ}-\eq{ENDEQ}.
\end{widetext}

 Before proceeding to the results we note that these solutions are the same as the steady-state solutions to the two-mode model of Refs.~\cite{Facchinetti,FacchinettiLong}, except that in these papers the counterparts of the coefficients $S_{xx}$ and $S_{zz}$ are primarily extracted from the from the eigenvalues of the time evolution of the finite lattice system. The authors then proceed to extrapolation to an infinite lattice and present results either the same or similar to ours, although there are differences in both the logic and the use of approximations (of which we have none). References~\cite{Facchinetti,FacchinettiLong} show no complicated polarizability tensors, and it might come as a surprise that the models are materially equivalent. The explanation is that the steady state is already built into the polarizability tensor. It is not valid  while electric field and polarization depend on time, whereas both time dependent relations between electric field and polarization and steady state could be solved from the differential equations in Refs.~\cite{Facchinetti,FacchinettiLong}. On the other hand, since the system is linear, frequency dependence of the steady state also uniquely determines the time dependent response, so we could go backward and find the time dependence if needed.
 
Putting in the incoming field $\bE_0 = \hat{\bf e}_x$, the radiated (far field) component of the electric field $ r = \hat{\bf e}_x\cdot\bE_T$ is directly the reflection amplitude.  We have
\beq
r = -\frac{3}{2\left[\delta +\hbox{$\frac{3}{2}$} S_{xx}+i-\displaystyle\frac{\omega _B^2}{\delta +\hbox{$\frac{3}{2}$} S_{zz}+ i}\right]}\bar T(a),
\label{LITTLER}
\eeq
where $\bar T(a)$ is given by~Eq.~\eq{TBARA} as before. The reflection coefficient $R$, as in Fig.~\ref{FACCFIG}, equals $R=|r|^2$.

The familiar structure of polarizability is emerging here, compare with Eq.~\eq{dFORM}. In the absence of the magnetic field, $\omega_B=0$, the result agrees with the one we already have for perpendicular incidence. The reflection amplitude can equal zero  only if the fractional expression in the denominator diverges at some detuning $\delta$, which, in turn, is only possible if $\Im[S_{zz}] = -\frac{2}{3}$. This, in fact, was the case for all values of $a\in(0,2\pi)$ we tried. For a given lattice spacing $a$, there is a detuning equal to $-\hbox{$\frac{3}{2}$} \Re[{\sf S}_{zz}]$ at which the term proportional to $\omega_B^2$ diverges, which gives $r=0$, and the lattice appears totally transparent. This holds no matter what the nonzero value of the magnetic field  is.

There is a physical explanation for the special value of the imaginary part of ${\sf S}_{zz}$. In the case of perpendicular incidence the self-sum tensor ${\sf S}$ has an eigenmode in the $z$ direction that cannot radiate. By energy conservation, this means that the mode is not damped either. The linewidth analogous to the linewidth for the transverse components of polarization would therefore be $\Gamma_z = 1 + \frac{3}{2}\Im[S_{zz}]=0$, giving $\Im[S_{zz}]=-\frac{2}{3}$. Energy conservation  together with the transverse nature of the electric field dictates the presence of subradiance, not cooperation between the atoms per se.

As we have noted already, the imaginary part of ${\sf S}_{xx}$ is also known. Let us write
\beq
{\sf S}_{xx} = R_{x} + \left(\frac{2\pi}{a^2}- \frac{2}{3}\right) i,\quad {\sf S}_{zz} = R_{z} -\frac{2}{3} i\,,
\label{IMPARTS}
\eeq
obviously with $R_x = \Re[S_{xx}]$, $R_z = \Re[S_{zz}]$. 
The reflection amplitude is bound by one in absolute value, and the maximum absolute value 1 is attained at two values of the detuning $\delta$ that depend on the real parts of the coefficients of the matrix $\sf S$ and on the magnetic field:
\beq
\delta_{\pm} =
 -\frac{3}{4}\left(
 R_x+R_z\pm
\sqrt{ \left(R_x-R_z\right)^2+\left(\hbox{$\frac{4}{3}$}\, \omega _B\right)^2}
\right).
\eeq
We then have both perfect reflection, and zero transmission.

We can calculate $R=|r|^2$, $T=|1+r|^2$, and $R+T$ from Eqs.~\eq{LITTLER} and~\eq{IMPARTS}, and ask if the result equals one for all detunings $\delta$ and magnetic-field splittings $\omega_B$. As another peculiar point of mathematics, this turns out to be the case as soon as the imaginary parts of ${\sf S}_{xx}$ and ${\sf S}_{zz}$, as well as $\bar{T}(a)$, all have their known values.  $R_x$ and $R_z$  have some definite values for each lattice constant, but energy conservation and the main qualitative features of the optical response do not depend on these values.

\begin{figure}
\includegraphics[width=0.9\columnwidth]{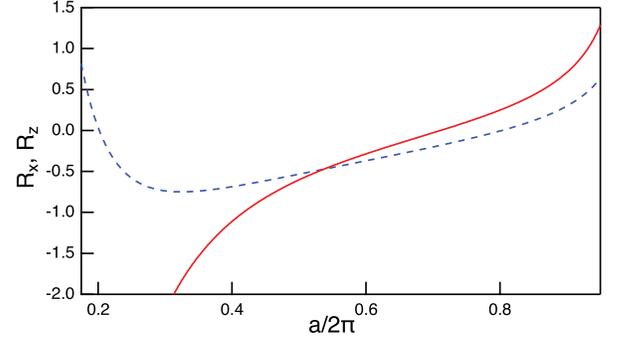}
 \caption{The real part of the elements  $S_{xx}$ (blue dashed line) and $S_{zz}$ (red solid line) of the self-sum tensor as a function of lattice spacing scaled to wavelength. This figure is for perpendicular incidence.}
\label{RESSHIFTS}
\end{figure}

The $z$ component of polarization may be found similarly to the radiated field. At the detuning when only the $z$ component remains, $\delta = -\frac{3}{2}\,R_z$, we have the dipole moment amplitude per atom
\beq
d_z = \frac{24 i\,\omega_B}{9 (R_x-R_z)^2+16 \omega_B^2}\,.
\eeq
This can be large, and especially so if the resonances associated with the in-plane and perpendicular dipoles coincide, $R_x=R_z$. We draw these resonance shifts in Fig.~\ref{RESSHIFTS} as a function of the lattice spacing, $R_x$ as a dashed blue line (this is part of the dashed blue line in Fig.~\ref{FS}) and $R_z$ as a  solid red  line.  For the lattice spacing of $0.537\times 2\pi$  (compare with the lattice spacing $0.55\times 2\pi$ in Fig.~\ref{FACCFIG}) such a confluence of the resonances in fact occurs.

\section{Discussion}\label{DISCUSSION}

We have investigated the optical response of atoms bound to a 2D lattice, allowing for non-normal incidence and anisotropic polarizability that occurs when a magnetic field is present. We have  also explicitly looked into the issues that come up when one stacks 2D lattices to make 3D lattice structures. Our methods are a mix of analytical and numerical calculations, symbolic computation, and numerical analysis. Especially the algebra associated with Sec.~\ref{MAGFIELD} is massively tedious, and might be difficult to get right without a tool such as Mathematica. Our amalgam of computer-based methods should be of some interest in its own right.

Of the physics principles that have come up, we emphasize energy conservation. In steady state an atom reradiates all of the energy that it removes from the electromagnetic field. With a plane wave coming in to a 2D lattice of atoms, in the absence of Bragg scattering, interference between the radiators dictates that only transmitted and reflected plane waves can be present in the far field. It is not a particular surprise that under such circumstances  the sum of reflection and transmission coefficients equals one in all cases we have tried. What boggles the mind is how the mathematics and numerics with the self-sum tensor, transfer matrix, and polarizability tensor always manage to line up in this way.

We turn the puzzle into a virtue, however. The self-sum and the transfer sum are not absolutely convergent, and by doing them differently could have given {\em very} different results. We have tamed the sum with an exponential convergence factor, an old and widely used trick everywhere in theoretical physics. The resulting $R+T=1$ lends some credence to this process.

While studying light transmission in a stack of 2D lattices, we have pointed out the close analogy of the system with atoms inside a 1D waveguide for light. We have argued that it is possible to make the stack to emulate light propagation in the 1D situation, but also discuss the limitations of such a scheme.

The original motivation for the present study was Ref.~\cite{Bettles_prl16} that headlines the observation from numerical simulations that a 2D lattice can be opaque. While one might be tempted to ascribe this behavior to cooperative effects, we have pointed out that the combination of energy conservation and single-Lorentzian form of the resonance line of the lattice suffices for this result. As a matter of principle, cooperativity does not enter the argument at all. Related discussions have appeared in the literature~\cite{Jenkins_2d_lat,Facchinetti,FacchinettiLong}, but we have also turned our argument into a tool to discuss the behavior of the lattice of atoms at non-normal incidence.

Indeed, building on the analyses of 1D vs.\ 2D lattices and  non-normal incidence, we have produced an example that might be relevant in experiments. We have shown that, by picking a suitable propagation direction and polarization for the driving light, it may be possible to cancel density dependent resonant shifts for atoms bound to a 3D lattice. This point is not trivial: The atoms, being constrained, do not move and collide, but the dipole-dipole interactions that are a major contribution to collision shifts still remain.  In general, density shifts are present even for trapped atoms.

Finally, we have developed quantitative theory of the optical response for a configuration with a magnetic field acting on a lattice of atoms.  This scheme was developed in Refs.~\cite{Facchinetti,FacchinettiLong} starting from finite-size numerical simulations. Besides giving semi-analytical results for an infinite lattice, we show that in this particular configuration any nonzero magnetic field can induce perfect transmission of light, and a total of two tunings of the light exist when the lattice is perfectly reflecting. The transparency can again be traced to back to energy conservation, but the total reflection does not match the single-resonance scheme we have identified for the case with no magnetic field. At present we do not understand the reason for perfect reflection, let alone how it gets embedded into the mathematics.

Our explicit solutions are seemingly limited to narrow special cases. A discussion of the restrictions is warranted.

With some lesser details such as no recoil effects and absence of optical pumping discussed elsewhere~\cite{JAV17}, the classical electrodynamics we have used is valid basically in the limit of low light intensity~\cite{Ruostekoski1997a}. It is probably impossible in practice to solve this problem in full quantum electrodynamics, and analyses that implicitly or explicitly assume that at any time at most a single photon is present invariably seem to come back to classical electrodynamics~\cite{PRA00,SVI10,Balik2013,Bienaime2013}. Even if the literature about lattice systems of atoms interacting with light is thick with quantum systems, with a few notable exceptions (e.g.,~\cite{TIE14,kimblemanybody}) both the atoms and  the light have so far mostly behaved essentially classically in both theories and experiments, and as such fit our framework.

The assumption about a 2D square lattice is benign; the same methodology works for any simple (Bravais) lattice. In the general case there will be three atomic resonances, and simplifications can be sought based on symmetries of the system. Lattices with an $n$-atom basis can be dealt with in an analogous way~\cite{PER17}, though the dimensions of the matrices would then be $3n\times3n$. Possible resonance frequencies multiply accordingly, and the $3n$-dimensional eigenvectors need not be transparent.  If the 2D Bravais lattices are not far enough apart, even for perpendicular incidence with $N$ lattices we have at least $N$ simultaneous equations to solve. That was, in fact, done for Figs.~\ref{TRMS}-\ref{EVFIG}. If the angle of incidence is not perpendicular, the number of coupled linear equations necessarily triples, and for an $n$-atom basis the dimensionality of the linear algebra problem becomes $3nN$. And so on. These are straightforward technical complications, although the numerical effort can grow and the hopes for qualitative understanding can dwindle drastically.

An infinite lattice is not possible in practice, but it is a feature not a bug in our theory. One can easily solve light propagation in atomic systems in all relevant dimensions with direct numerical simulations. However, the demands on computational resources would usually dictate a limited number of the atoms, and a limited size of the sample. The optics of finite-size samples would then have to be disentangled from the simulation results~\cite{JAV17}. We wish to unearth generic principles, and it would get difficult. In the same vein, we do not produce listings of quantitative results, but instead we have built a toolbox that anyone interested in specific questions in this area might find helpful.

\begin{acknowledgments}
This work is supported in part by the National Science Foundation, Grant Nr. PHY-1401151.
\end{acknowledgments}

\appendix

\section{Numerical details}\label{NUMDET}
\subsection{Richardson extrapolation}\label{NUMDET-R}
In the computations of the self-sum and the transfer sum we have to do two limits. We wish to have automated control of the limits, given a prescribed goal for precision. This is a fairly demanding problem in numerical analysis. Here we describe our solution.

By symmetry, the sum in~\eq{SELFSUM} can be carried out in the first quadrant with only nonnegative integers, which saves 75\% of the numerical effort.
 
Next, consider the sum in Eq.~\eq{SELFSUM}  for a given $\eta$. We would like to know how large a value of the upper limit $M=M(\eta)$ is needed to reach a prescribed relative error $\epsilon_M$ from the $M\rightarrow\infty$ limit of the sum. To this end, let us estimate the residual by replacing the discrete sum by an integral, using the leading $1/n$ dependence of the summand in the estimate. We find
\beq
R_S(M) = \int_{n\ge M} d^2n\, \frac{e^{(i -\eta)na}}{2 na} = \frac{\pi  e^{-a (\eta -i) M}}{a^2 (\eta -i)}\,.
\eeq
In the limit $M\rightarrow0$ the integral does not represent the infinite sum particularly well  for several obvious reasons, but it does give an idea of the scale of the sum
\beq
R_S(0) =  \frac{\pi  }{a^2 (\eta -i)}\,.
\label{ESTIMATE}
\eeq
The requirement that the relative error of the sum for the given upper limit $M$ equals $\epsilon_M$ gives the equation $|R_S(M)/R_S(0)| = \epsilon_M$, from which we may solve the upper limit $M$:
\beq
M(a,\eta) = -\frac{1}{a\eta} \ln\epsilon_M.
\label{SELFM}
\eeq
We add another heuristic condition that the value $M(a,\eta)$ be at least 16.780913. The essentially random non-integer value is an attempt to reduce the probability that including or excluding a member $\bf n$ in the sum~\eq{SELFSUM} depends on round-off errors.

One may study the precision by which the sum for any given upper limit $M$ approximates the infinite sum by making $M$ larger, and seeing how much the result changes. In this way we have found empirically that the expression~\eq{SELFM} for the limit $M$ is numerically useful when one aims at a relative truncation error of the sum on the order of $\epsilon_M$. The computation time scales like $M^2$, and thus like $1/a^2$ for small $a$. We have devised various ad-hoc procedures to mitigate this problem, but most expediently we avoid small values of $a$ that apparently do not yield interesting results anyway.

The final limit $\eta\rightarrow0+$ is more tricky.  As may be seen from Eqs.~\eq{SELFSUM} and~\eq{SELFM}, the number of the terms in the sum scales as $1/\eta^2$,  and the numerical effort can get substantial in a brute-force attempt to make $\eta\rightarrow0$.

We resort to Richardson extrapolation \cite{NUMRES} instead. First we calculate a sequence of values for the sum, decreasing the convergence parameter $\eta$ by a factor $2$ every time:
\beq
S^{(0)}_n = S(a,\eta/2^n),\quad n=0,1,2,\ldots\,.
\label{RSEQ}
\eeq
Assuming an expansion of the form
\beq
S(a,\eta) = S(a) + K_1 \eta + K_2 \eta^2 + \ldots,
\label{ERROREXPANSION}
\eeq
for the sum as a function of the parameter $\eta$,
we see that for small enough $\eta$ the ratio
\beq
\frac{S^{(0)}_{n}-S^{(0)}_{n+1}}{S^{(0)}_{n+1}-S^{(0)}_{n+2}}
\label{RATIO}
\eeq
should be approximately 2 for all $n$. Empirically, it is so for all but some isolated values of $a$. So, next we form a sequence of numbers
\beq
S^{(1)}_n ={2 S^{(0)}_{n+1} - S^{(0)}_{n}}
\eeq
that has the property that the terms proportional to $\eta$ cancel, and the leading error in the sequence $S^{(1)}_n$  is proportional to $(\eta/2^n)^2$. This would mean that the ratio as in Eq.~\eq{RATIO} for the sequence $S_n^{(1)}$ {should be approximately 4, which it empirically is. Continuing iteratively in this way, the leading error in the sequence 
\beq
S^{(k)}_n = \frac{2^k S^{(k-1)}_{n+1} -S^{(k-1)}_{n}}{2^k-1},\quad k=1,2,\ldots,
\eeq
should be proportional to $(\eta/2^n)^{k+1}$. One can go in this way to higher orders $k$ to accelerate the convergence.

The main numerical issue is that the limit $\eta\rightarrow0$ apparently genuinely diverges at the onset of new Bragg reflections, like $\eta^{-1/2}$ at $a=2\pi$. As far as we can tell, the sum converges for all values of $a$ that do not give a Bragg reflection in the plane of the lattice, but when $a$ approaches a new Bragg reflection, the values of the parameter $\eta$ for which the expansion~\eq{ERROREXPANSION} is useful become smaller and smaller. Correspondingly, Richardson extrapolation in the order $k$ may seem to converge, but gives a spurious result.

Our method to combat this problem runs as follows. We compute members of the sequence~\eq{RSEQ} one by one, typically starting with $\eta=0.08$ that has proven convenient for our purposes. Once we have three values, we do a Richardson extrapolation to the order that we have denoted by $k=2$, with an error proportional to $\eta^3$. We then keep adding terms to the sequence~\eq{RSEQ} and do the Richardson extrapolation with the {\em last three members\/} of the sequence to the order $k=2$, until the result is either deemed convergent or the process takes too much time and is terminated.

Specifically, if Richardson extrapolation works as advertised, the result   with the leading error from using the last three terms of the sequence~\eq{RSEQ} up to the order $n$ is
\beq
{\tilde S}_n = S(a) + K \left(\frac{\eta}{2^{n}}\right)^3\,,
\eeq
where $K$ is approximately a constant. Given this expression, we may then combine the last two extrapolation results into the formulas
\bea
S(a) &=& \frac{8 \tilde S_n-\tilde S_{n-1}}{7}\,.\label{REXRES}\\
\frac{\tilde S_n -S(a) }{S(a)}&=&- \frac{\tilde S_n-\tilde S_{n-1}}{8\tilde S_n-\tilde S_{n-1}}\,.
\eea
The former is in fact the Richardson extrapolation to the next order $k=3$ with the error $\propto \eta^4$, and the latter is essentially an estimate for the relative error of the extrapolation to the order $k=2$.  We terminate the sequence~\eq{RSEQ} at the first sequence order $n$ such that, for the given relative error $\epsilon$, we have
\beq
\frac{|\tilde S_n-\tilde S_{n-1}|}{|8\tilde S_n-\tilde S_{n-1}|}\ \le\epsilon\,,
\label{CONVCRIT}
\eeq
and declare the result~\eq{REXRES}.
If convergence does not occur by some order $n$, typically 8, we call the result ``error.'' Of course, by increasing the limit order $n$, we could in principle force convergence whenever we are not at an exact Bragg diffraction value of $a$.

The final caveat is that the numbers $S^{(0)}_n$ obtained by truncating the sums at the limiting values $M(\eta)$ are already approximate, aiming at a relative error of $\epsilon_M$. If the noise in the sequence $S^{(0)}_n$ from this approximation is too large, Richardson extrapolation will not converge to the accuracy $\epsilon$. We safeguard against this failure by choosing $\epsilon_M = 0.1\epsilon$.  In our numerical surveys we typically use the relative-error parameter $\epsilon = 10^{-3}$. Except very close to Bragg reflection points, this indeed gives relative errors smaller than or comparable to $\sim 10^{-3}$. Such errors are barely noticeable by eye in figures such as Fig.~\ref{FRT}. 

The same process may be used for the transfer sum $T(z,a)$. For the transfer sum the limit $\eta\rightarrow0$ apparently also diverges at Bragg reflection points such as $a=2\pi$, and presents numerical problems nearby.

The basically same numerics also works for the tensor counterparts of the self-sum and transfer sum, $\sf S$ and $\sf T$, except that it deals with $3\times3$ complex matrices not complex numbers. In the convergence criterion~\eq{CONVCRIT} we use the metric induced by the $\ell^2$ norm of a matrix so that, for instance, the ``length'' of the matrix $\sf S$ equals\
\beq
|{\sf S}|=\sqrt{\sum_{i,j=1}^3 |{\sf S}_{ij}|^2}\,.
\eeq
For numbers, $1\times1$ matrices, this concept boils down to the absolute value as in~\eq{CONVCRIT}, and in the matrix case it is not thrown off by very small entries in the matrices that occur because round-off errors cause nonzero values where the exact result would be zero.

We were able to run all of the results we needed with controlled precision as per above, on Mathematica for the scalar case and using mostly C++ for the matrix case. We can think of significant improvements to the numerical procedures, but the return would probably not justify the effort.

\begin{widetext}
\subsection{Transfer matrix}\label{NUMDET-T}
As to the long-distance analytical form of the transfer matrix,  \eq{MATRICES} and~\eq{ROTMAT}, we firstly without further ado replace the sum~\eq{TRFMATRIX} with an integral over the site index ${\bf n}$, and express it in terms of polar coordinates in the plane. The integral over the angles is easy to carry out analytically using Mathematica. We choose the propagation vector in Eq.~\eq{KVECTOR} with $\phi=0$, and are left with integrals such as
\beq
\frac{a^2}{2\pi  i} {\sf T}_{xz}\! =\!\!  \int_0^\infty\!\!\!\! d\rho\,\frac{e^{-\eta\rho + i(\sqrt{z^2+\rho^2}-z \cos\theta)}
z\rho^2\left(\!-3 z^2+\rho^2+3i\sqrt{z^2+\rho^2}\right)\!J_1(\rho\sin\theta )
}{(z^2+\rho^2)^{5/2}},
\eeq
where $J$ stands for the Bessel function. We do this integral numerically, again using Mathematica. For very small $\eta>0$ the value turns out the be independent of $z>0$, and by varying the angle $\theta$ we find that it equals $-\sin\theta$. This is the upper-right element of the matrix $M(\theta)$ in Eq.~\eq{MATRICES}. The other elements come from similar arguments. Given the simple results it is obvious that a completely analytical procedure to find the elements of the matrix $M(\theta)$ should exist, but we have not bothered to find one. Finally, the matrix $R(\phi)$ of Eq.~\eq{ROTMAT} represents rotation about the $z$ axis, and serves to extend the result to an arbitrary azimuthal angle $\phi$ under the evidently correct assumption that in such rotations of the direction of the incoming light the matrix $\sf T$ transforms as a tensor.
\end{widetext}

\section{Bragg scattering}\label{BRAGREF}
Like in the main text, let us denote the wave vector of the incoming light by
\beq
\bk = \sin\theta \cos\phi\, \hat{\bf e}_x +  \sin\theta \sin\phi\, \hat{\bf e}_y +  \cos\theta\, \hat{\bf e}_z\,.
\eeq
Its projection to the plane of the atoms is
\beq
\bk_\parallel =  \sin\theta \cos\phi\, \hat{\bf e}_x +  \sin\theta \sin\phi\, \hat{\bf e}_y\,.
\eeq

The incoming light paints a phase pattern on the atomic dipoles, which become proportional to $e^{i\bk_\parallel\cdot {\bf R}_{\bf n}}$. The question is about the spatial pattern, especially in the far field, that the dipoles radiate in return. More specifically, there will be at least two obvious plane waves, the reflected field and the radiated part of the transmitted field. Our goal is to characterize the wave vectors $\bar\bk$ for all possible radiated plane waves.

The principle that cracks the case is that a plane wave with the wave vector $\bar\bk$, when incident on the atoms, would have to paint the same phase pattern $e^{i\bk_\parallel\cdot {\bf R}_{\bf n}}$. Any vector $\bar\bk$ whose projection to the plane satisfies $\bar\bk_\parallel = \bk_\parallel$ fits the bill. However, not all of the $\bar\bk$ with this property qualify. Namely, in steady state the frequency of the light is fixed, and therefore so is the absolute value of the vector $\bar\bk$; $|\bar\bk|=1$ in our units.

These considerations immediately leave two alternatives: a wave with $\bar\bk = \bk$ that co-propagates with the incoming plane wave, and the reflected wave $\bar\bk=\bk_R$ in which the $z$ component has flipped the sign. The transmitted wave is the superposition of the incoming field and the reradiated field with the wave number $\bk$.

Nonetheless, we have not yet exhausted all possibilities. The dipoles reside at discrete positions, and any $\bar\bk_\parallel$ that gives the same phase pattern as $\bk_\parallel$ at the positions of the dipoles also qualify. This occurs precisely when these vectors differ by a reciprocal lattice vector,
\beq
\bar\bk_\parallel = \bk_\parallel +{\bf K},\quad {\bf K} = \frac{2\pi}{a}(m_x\hat{\bf e}_x+m_y\hat{\bf e}_y),
\eeq
where $m_x$ and $m_y$ are arbitrary integers. But the radiation still has to occur at a wave vector whose absolute value equals $1$, so we have the condition ${\bar\bk_\parallel}^2\le 1$ to leave room for the $z$ components $\bar{k}_z = \pm\sqrt{1-{\bar\bk_\parallel}^2}$. We have an inequality that may be rearranged to give
\beq
\frac{4 \pi ^2\!\! \left(m_x^2\!+\!m_y^2\right)}{a^2}+\frac{4 \pi  \sin \theta 
   (m_x\! \cos \phi \!+\!m_y\! \sin \phi )}{a}+\sin ^2(\theta ) \le 1.
\label{BRAGC}
\eeq

Here $m_x=m_y=0$ always qualify, and give the forward and backward scattered waves. Any other combination of $m_x$ and $m_y$ that satisfy the inequality~\eq{BRAGC} signals the presence of a nontrivial Bragg-scattered wave. The condition for Bragg scattering depends on the azimuthal angle $\phi$. If this angle is varied freely, Bragg scattering for a given order $m_x,\,m_y$ may be found for lattice spacings that satisfy
\beq
a \ge \frac{2\pi\sqrt{m_x^2+m_y^2}}{1+\sin\theta}\,.
\eeq
Bragg scattering may thus occur for $a>\pi$, starting from a half-wavelength lattice spacing when the angle of incidence gets extremely oblique, $\theta\rightarrow\pi/2$. On the other hand, by having the incoming wave propagate at an angle $\phi=\pi/4$ so that the projection $\bk_\parallel$ bisects the angle between lattice axes, one may delay the onset of Bragg scattering up to $a=\sqrt2\, \pi$.

2D Bragg scattering is also valid Bragg scattering in an evenly spaced 3D stack of 2D lattices. One may show easily that if there is no 2D Bragg scattering in a square lattice, there will be no nontrivial 3D Bragg scattering in a simple cubic lattice made by stacking 2D square lattices either. Other than warning that this may change if the spacing between the 2D lattices is larger than the lattice constant, we do not go into the details.

\section{The mean-field approximation}
\label{MFATEST}

Our mean-field approximation for radiation in the lattice plane states that, if the fraction of lattice sites occupied is $\zeta$, on the average the field radiated by the other atoms on any given atom, the self-sum $S$, and the shift of the resonance $s=-\frac{3}{2}\Re[S]$, all get multiplied by $\zeta$. In this Appendix~\ref{MFATEST} we demonstrate with specific examples that the approximation is reasonable.

The mean-field approximation is not trivial because of multiple scattering and the ensuing cooperative phenomena: An atom radiates a field, which strikes another atom, which radiates a field, which comes back to the original atom, and so on. Here we study cooperative effects in the spirit of our earlier simulations of light propagation in an atomic sample~\cite{Javanainen2014a,JAV17}, by solving a set of linear equations that takes into account both the incoming light and the secondary fields from all atoms. The solution  will give, among other things, the total field from all other atoms falling on any given atom.

In our  examples we take the sample to be a circular cut from the lattice with the spacing $a=\pi$ (half of a wavelength). We include the sites that are at most the distance $R$ from the origin\commentout{, a total of 5025 of them}. Each site is occupied by an atom with the probability $\zeta$. We take a circularly polarized plane wave with the amplitude $E_0$ coming in on the lattice in a perpendicular direction, on resonance so that $\delta=0$.

If there are empty sites,  even at the center of the sample the radiation from the atoms in general does not have the same circular polarization as the incoming light. However, on the average the polarization  would be circular at the center, and in an infinite lattice at any site. We  therefore project the electric field to the polarization of the incoming light, and characterize the secondary radiation from the other atoms on a given atom with a single complex amplitude. We compute this electric field $\bar E$ either at the center of the sample, or at a lattice site closest to the center that is occupied.

Modeling the secondary field as $\bar E= d S$, one can easily solve from Eqs.~\eq{TOTALFIELD} and~\eq{dFORM} the self-sum $S$, and obtain the shift of the resonance:
\beq
s = \Re\left[\frac{(\delta+i)\bar E}{\bar E+E_0}\right].
\eeq
We find it for a number of samples with the sites filled at random with the probability $\zeta$, and report the average as well as its standard deviation deduced from the samples. The fluctuations from sample to sample can be large, so the number of samples used in the averaging range up to a million. The computations are implemented by deriving suitable C++ classes from the ones used in our earlier 3D simulations, and the code works internally in 3D. Our standard sample radius $R=40\, a$ encloses $5025$ lattice sites, which means that the key operation is to solve up to a $15075\times 15075$ set of dense linear equations.

In the case with $R=40 \,a$ and $\zeta=1$, all sites occupied, the result for the finite-size sample is $s = 0.7958$. It should be compared with the result for an infinite lattice from Sec.~\ref{SELFSUMCOMP}, $s = 0.8006$. This level of the agreement, $0.6\%$, is probably fortuitous, but it lends credence to our methods.

\begin{figure}
\hspace{-30pt}\includegraphics[width=0.9\columnwidth]{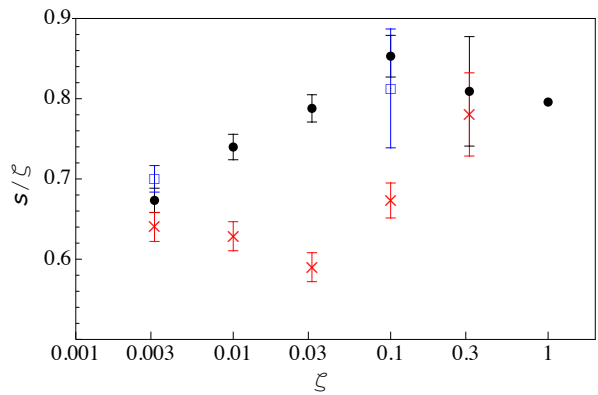}
 \caption{Line shift $s$ divided by the filling factor $\zeta$ as a function of the filling factor $\zeta$, with statistical standard deviations. Black filled circles: $R=40\,a$, shift for an atom closest to the center of the sample. Red crosses: $R=40\,a$, shift for an atom that would reside at the center of the sample, whether there is an atom or not. Blue open boxes: $R=80\,a$, for an atom closest to the center of the sample. }
\label{RESSHIFTS}
\end{figure}

Representative results are shown in Fig.~\ref{RESSHIFTS}, which plots the shift divided by the filling factor, ${s/\zeta}$, as a function of $\zeta$ over a range of two and a half orders of magnitude. The standard deviations are also given. Black filled circles are for the sample radius $R=40 \,a$, and the field is computed at an occupied lattice site closest to the center. Data marked by red crosses are computed at the center of the circuclar cut of the lattice, whether there is an atom at the center or not. The points marked by blue open squares are for  $R=80 \,a$, and for an atom closest to the center.

In an infinite lattice, if it did not matter whether we are looking at a filled or non-filled site, and if the mean-field approximation were exact, the results would all come out at the same value of $s/\zeta$.
The range of the values of  $s/\zeta$ in the figure is on the order of a few tens of per cent.  Variation of the magnitude we are seeing in Fig.~\ref{RESSHIFTS} would not invalidate the conclusions of Sec.~\ref{CANCELLIGHTSHIFTS}. Besides, much of it evidently comes from  factors other than the inaccuracy of the mean field approximation per se.

\section{Magnetic field and polarizability}\label{POLAPP}
In the present Appendix~\ref{POLAPP} we derive the polarizability tensor of an atom with a $J=0\rightarrow J=1$ transition in the presence of the magnetic field specified in terms of the polar angles $\theta$ and $\phi$ as
\beq
{\bf B} =  B(\sin\theta \cos\phi\, \hat{\bf e}_x +  \sin\theta \sin\phi\, \hat{\bf e}_y +  \cos\theta\, \hat{\bf e}_z)
\eeq
{\em classically}. We think of the atom as a charged ($q$) damped ($\gamma$) harmonic oscillator (frequency $\omega_0$, mass $m$) under the influence of both the driving electric field $\bE(t) = \half\bE e^{-i\omega t} + \rm{c.c.}$ and the constant magnetic field {\bf B}.

Newton's second law gives the equation of motion for the position of the charge
\beq
m\ddot\br = -m\omega_0^2 \br^2 - 2 m\gamma\dot\br +  \half q\bE e^{-i\omega t} + q \dot\br\times{\bf B}.
\eeq 
We look for the steady state with the Ansatz $\br(t) = \half \br_0 e^{-i\omega t}$. The true physical solution is $\br(t) + \br^*(t)$, so that $\bd=q\br_0$ is the positive-frequency component of the dipole moment resulting from the driving field with the positive-frequency component $\bE$. The dipole moment is a linear function of the electric field, $\bd = {\sf A}\bE$, which defines the polarizability tensor $\sf A$.

The calculations are tedious, so we proceed with just an outline that we have filled in with Mathematica. Like in the undergraduate exercise of a driven damped harmonic oscillator, we find resonance denominators of the type $\omega^2-\omega_0^2$, yet we know from modeling of resonance behavior of atoms that it is the detuning $\Delta = \omega-\omega_0$ that matters most. We therefore develop a ``resonance approximation'' as follows:
\begin{itemize}
\item[1.] Write the polarizability tensor in the form ${\sf A} = {\cal A}/\omega$. Up to the point when all results are put together, we deal with the tensor $\cal A$. Compare this with the standard undergraduate exercise, in which one would make the resonance approximation by writing the resonance denominator as $\omega^2-\omega_0^2 = (\omega-\omega_0)(\omega+\omega_0)\simeq 2\omega\Delta$. We are getting ahead of things here, but wish to point out that this scaling is particularly useful as in our custom units $\omega=1$.
\item[2.] Write the tensor $\cal A$ in terms of $\omega_0$ and $\omega=\omega_0+\Delta$.
\item[3.] Expand all components of the tensor $\cal A$ as partial fractions with the first power of $\Delta$ in the denominators, zeroth power in the numerators.
\item[4.] Replace each partial fraction with the leading term in its expansion in the limit $\omega_0\rightarrow\infty$.
\item[5.] Keep only the partial fractions that are of the dominant, zeroth, order in $\omega_0$.
\end{itemize}

This approximation produces terms with resonance denominators of the form $\Delta + i \gamma$ and $\Delta\pm \Omega_B + i\gamma$, where $\Omega_B ={qB}/{2m}$ is {\em half\/} of the cyclotron frequency for the given magnetic field $B$. Finally, we compare these results in the case of zero magnetic field with the expression~\eq{POLARIZABILITY} for scalar polarizability, and write the corresponding expressions with the resonance denominators that include the magnetic field. The polarizability tensor is unwieldy, but here is the $xx$ component as an example:
\begin{widetext}
\bea
A_{xx}&=& -\frac{3 \{\cos [2 (\theta -\phi )]+\cos [2 (\theta +\phi )]+2 \cos (2 \theta )-2 \cos (2
   \phi)+6\}}{32 \left(\delta -\omega _B+i\right)}\nonumber\\
   && -\frac{3 \{\cos [2 (\theta -\phi )]+\cos [2 (\theta +\phi )]+2 \cos (2 \theta )-2 \cos (2
   \phi)+6\}}{32 \left(\delta+\omega _B+i\right)}\nonumber\\
   &&-\frac{3 \sin ^2(\theta ) \cos ^2(\phi )}{2 (\delta +i)}\,,
 \eea
with $\omega_B=\Omega_B/\gamma$.
\end{widetext}

While the derivation we have outlined is classical and seems ad-hoc, the results agree with fully quantum mechanical numerical calculations as reported by the ``Software Atom''~\cite{JAV17SA} for the $J=0$ to $J'=1$ transition. In the quantum case $\omega_B$ equals the frequency difference caused by the magnetic field between the states $m'=1$ and $m'=0$ of the excited-state manifold, expressed in units of the linewidth of the transition, and is negative if the Lande factor for the level $J'=1$ is positive. Another point in favor of our version of the resonance approximation is that energy is conserved: We have checked with an explicit calculation that, in steady state and over a period of the oscillations, the energy that the driving electric field puts on the charge equals the energy that the oscillating dipole radiates.


\begin{thebibliography}{51}%
\makeatletter
\providecommand \@ifxundefined [1]{%
 \@ifx{#1\undefined}
}%
\providecommand \@ifnum [1]{%
 \ifnum #1\expandafter \@firstoftwo
 \else \expandafter \@secondoftwo
 \fi
}%
\providecommand \@ifx [1]{%
 \ifx #1\expandafter \@firstoftwo
 \else \expandafter \@secondoftwo
 \fi
}%
\providecommand \natexlab [1]{#1}%
\providecommand \enquote  [1]{``#1''}%
\providecommand \bibnamefont  [1]{#1}%
\providecommand \bibfnamefont [1]{#1}%
\providecommand \citenamefont [1]{#1}%
\providecommand \href@noop [0]{\@secondoftwo}%
\providecommand \href [0]{\begingroup \@sanitize@url \@href}%
\providecommand \@href[1]{\@@startlink{#1}\@@href}%
\providecommand \@@href[1]{\endgroup#1\@@endlink}%
\providecommand \@sanitize@url [0]{\catcode `\\12\catcode `\$12\catcode
  `\&12\catcode `\#12\catcode `\^12\catcode `\_12\catcode `\%12\relax}%
\providecommand \@@startlink[1]{}%
\providecommand \@@endlink[0]{}%
\providecommand \url  [0]{\begingroup\@sanitize@url \@url }%
\providecommand \@url [1]{\endgroup\@href {#1}{\urlprefix }}%
\providecommand \urlprefix  [0]{URL }%
\providecommand \Eprint [0]{\href }%
\providecommand \doibase [0]{http://dx.doi.org/}%
\providecommand \selectlanguage [0]{\@gobble}%
\providecommand \bibinfo  [0]{\@secondoftwo}%
\providecommand \bibfield  [0]{\@secondoftwo}%
\providecommand \translation [1]{[#1]}%
\providecommand \BibitemOpen [0]{}%
\providecommand \bibitemStop [0]{}%
\providecommand \bibitemNoStop [0]{.\EOS\space}%
\providecommand \EOS [0]{\spacefactor3000\relax}%
\providecommand \BibitemShut  [1]{\csname bibitem#1\endcsname}%
\let\auto@bib@innerbib\@empty
\bibitem [{\citenamefont {Keaveney}\ \emph {et~al.}(2012)\citenamefont
  {Keaveney}, \citenamefont {Sargsyan}, \citenamefont {Krohn}, \citenamefont
  {Hughes}, \citenamefont {Sarkisyan},\ and\ \citenamefont
  {Adams}}]{Keaveney2012}%
  \BibitemOpen
  \bibfield  {author} {\bibinfo {author} {\bibfnamefont {J.}~\bibnamefont
  {Keaveney}}, \bibinfo {author} {\bibfnamefont {A.}~\bibnamefont {Sargsyan}},
  \bibinfo {author} {\bibfnamefont {U.}~\bibnamefont {Krohn}}, \bibinfo
  {author} {\bibfnamefont {I.~G.}\ \bibnamefont {Hughes}}, \bibinfo {author}
  {\bibfnamefont {D.}~\bibnamefont {Sarkisyan}}, \ and\ \bibinfo {author}
  {\bibfnamefont {C.~S.}\ \bibnamefont {Adams}},\ }\bibfield  {title} {\enquote
  {\bibinfo {title} {Cooperative {Lamb} shift in an atomic vapor layer of
  nanometer thickness},}\ }\href@noop {} {\bibfield  {journal} {\bibinfo
  {journal} {Phys. Rev. Lett.}\ }\textbf {\bibinfo {volume} {108}},\ \bibinfo
  {pages} {173601} (\bibinfo {year} {2012})}\BibitemShut {NoStop}%
\bibitem [{\citenamefont {Pellegrino}\ \emph {et~al.}(2014)\citenamefont
  {Pellegrino}, \citenamefont {Bourgain}, \citenamefont {Jennewein},
  \citenamefont {Sortais}, \citenamefont {Browaeys}, \citenamefont {Jenkins},\
  and\ \citenamefont {Ruostekoski}}]{Pellegrino2014a}%
  \BibitemOpen
  \bibfield  {author} {\bibinfo {author} {\bibfnamefont {J.}~\bibnamefont
  {Pellegrino}}, \bibinfo {author} {\bibfnamefont {R.}~\bibnamefont
  {Bourgain}}, \bibinfo {author} {\bibfnamefont {S.}~\bibnamefont {Jennewein}},
  \bibinfo {author} {\bibfnamefont {Y.~R.~P.}\ \bibnamefont {Sortais}},
  \bibinfo {author} {\bibfnamefont {A.}~\bibnamefont {Browaeys}}, \bibinfo
  {author} {\bibfnamefont {S.~D.}\ \bibnamefont {Jenkins}}, \ and\ \bibinfo
  {author} {\bibfnamefont {J.}~\bibnamefont {Ruostekoski}},\ }\bibfield
  {title} {\enquote {\bibinfo {title} {Observation of suppression of light
  scattering induced by dipole-dipole interactions in a cold-atom ensemble},}\
  }\href {\doibase 10.1103/PhysRevLett.113.133602} {\bibfield  {journal}
  {\bibinfo  {journal} {Phys. Rev. Lett.}\ }\textbf {\bibinfo {volume} {113}},\
  \bibinfo {pages} {133602} (\bibinfo {year} {2014})}\BibitemShut {NoStop}%
\bibitem [{\citenamefont {Kwong}\ \emph {et~al.}(2015)\citenamefont {Kwong},
  \citenamefont {Yang}, \citenamefont {Delande}, \citenamefont {Pierrat},\ and\
  \citenamefont {Wilkowski}}]{wilkowski2}%
  \BibitemOpen
  \bibfield  {author} {\bibinfo {author} {\bibfnamefont {C.~C.}\ \bibnamefont
  {Kwong}}, \bibinfo {author} {\bibfnamefont {T.}~\bibnamefont {Yang}},
  \bibinfo {author} {\bibfnamefont {D.}~\bibnamefont {Delande}}, \bibinfo
  {author} {\bibfnamefont {R.}~\bibnamefont {Pierrat}}, \ and\ \bibinfo
  {author} {\bibfnamefont {D.}~\bibnamefont {Wilkowski}},\ }\bibfield  {title}
  {\enquote {\bibinfo {title} {Cooperative emission of a pulse train in an
  optically thick scattering medium},}\ }\href {\doibase
  10.1103/PhysRevLett.115.223601} {\bibfield  {journal} {\bibinfo  {journal}
  {Phys. Rev. Lett.}\ }\textbf {\bibinfo {volume} {115}},\ \bibinfo {pages}
  {223601} (\bibinfo {year} {2015})}\BibitemShut {NoStop}%
\bibitem [{\citenamefont {Bromley}\ \emph {et~al.}(2016)\citenamefont
  {Bromley}, \citenamefont {Zhu}, \citenamefont {Bishof}, \citenamefont
  {Zhang}, \citenamefont {Bothwell}, \citenamefont {Schachenmayer},
  \citenamefont {Nicholson}, \citenamefont {Kaiser}, \citenamefont {Yelin},
  \citenamefont {Lukin}, \citenamefont {Rey},\ and\ \citenamefont
  {Ye}}]{Ye2016}%
  \BibitemOpen
  \bibfield  {author} {\bibinfo {author} {\bibfnamefont {S.~L.}\ \bibnamefont
  {Bromley}}, \bibinfo {author} {\bibfnamefont {B.}~\bibnamefont {Zhu}},
  \bibinfo {author} {\bibfnamefont {M.}~\bibnamefont {Bishof}}, \bibinfo
  {author} {\bibfnamefont {X.}~\bibnamefont {Zhang}}, \bibinfo {author}
  {\bibfnamefont {T.}~\bibnamefont {Bothwell}}, \bibinfo {author}
  {\bibfnamefont {J.}~\bibnamefont {Schachenmayer}}, \bibinfo {author}
  {\bibfnamefont {T.~L.}\ \bibnamefont {Nicholson}}, \bibinfo {author}
  {\bibfnamefont {R.}~\bibnamefont {Kaiser}}, \bibinfo {author} {\bibfnamefont
  {S.~F.}\ \bibnamefont {Yelin}}, \bibinfo {author} {\bibfnamefont {M.~D.}\
  \bibnamefont {Lukin}}, \bibinfo {author} {\bibfnamefont {A.~M.}\ \bibnamefont
  {Rey}}, \ and\ \bibinfo {author} {\bibfnamefont {J.}~\bibnamefont {Ye}},\
  }\bibfield  {title} {\enquote {\bibinfo {title} {Collective atomic scattering
  and motional effects in a dense coherent medium},}\ }\href
  {http://dx.doi.org/10.1038/ncomms11039} {\bibfield  {journal} {\bibinfo
  {journal} {Nat Commun}\ }\textbf {\bibinfo {volume} {7}},\ \bibinfo {pages}
  {11039} (\bibinfo {year} {2016})}\BibitemShut {NoStop}%
\bibitem [{\citenamefont {Guerin}\ \emph {et~al.}(2016)\citenamefont {Guerin},
  \citenamefont {Ara\'ujo},\ and\ \citenamefont {Kaiser}}]{Guerin_subr16}%
  \BibitemOpen
  \bibfield  {author} {\bibinfo {author} {\bibfnamefont {William}\ \bibnamefont
  {Guerin}}, \bibinfo {author} {\bibfnamefont {Michelle~O.}\ \bibnamefont
  {Ara\'ujo}}, \ and\ \bibinfo {author} {\bibfnamefont {Robin}\ \bibnamefont
  {Kaiser}},\ }\bibfield  {title} {\enquote {\bibinfo {title} {Subradiance in a
  large cloud of cold atoms},}\ }\href {\doibase
  10.1103/PhysRevLett.116.083601} {\bibfield  {journal} {\bibinfo  {journal}
  {Phys. Rev. Lett.}\ }\textbf {\bibinfo {volume} {116}},\ \bibinfo {pages}
  {083601} (\bibinfo {year} {2016})}\BibitemShut {NoStop}%
\bibitem [{\citenamefont {Jennewein}\ \emph {et~al.}(2016)\citenamefont
  {Jennewein}, \citenamefont {Besbes}, \citenamefont {Schilder}, \citenamefont
  {Jenkins}, \citenamefont {Sauvan}, \citenamefont {Ruostekoski}, \citenamefont
  {Greffet}, \citenamefont {Sortais},\ and\ \citenamefont
  {Browaeys}}]{Jennewein_trans}%
  \BibitemOpen
  \bibfield  {author} {\bibinfo {author} {\bibfnamefont {S.}~\bibnamefont
  {Jennewein}}, \bibinfo {author} {\bibfnamefont {M.}~\bibnamefont {Besbes}},
  \bibinfo {author} {\bibfnamefont {N.~J.}\ \bibnamefont {Schilder}}, \bibinfo
  {author} {\bibfnamefont {S.~D.}\ \bibnamefont {Jenkins}}, \bibinfo {author}
  {\bibfnamefont {C.}~\bibnamefont {Sauvan}}, \bibinfo {author} {\bibfnamefont
  {J.}~\bibnamefont {Ruostekoski}}, \bibinfo {author} {\bibfnamefont {J.-J.}\
  \bibnamefont {Greffet}}, \bibinfo {author} {\bibfnamefont {Y.~R.~P.}\
  \bibnamefont {Sortais}}, \ and\ \bibinfo {author} {\bibfnamefont
  {A.}~\bibnamefont {Browaeys}},\ }\bibfield  {title} {\enquote {\bibinfo
  {title} {Coherent scattering of near-resonant light by a dense microscopic
  cold atomic cloud},}\ }\href {\doibase 10.1103/PhysRevLett.116.233601}
  {\bibfield  {journal} {\bibinfo  {journal} {Phys. Rev. Lett.}\ }\textbf
  {\bibinfo {volume} {116}},\ \bibinfo {pages} {233601} (\bibinfo {year}
  {2016})}\BibitemShut {NoStop}%
\bibitem [{\citenamefont {Roof}\ \emph {et~al.}(2016)\citenamefont {Roof},
  \citenamefont {Kemp}, \citenamefont {Havey},\ and\ \citenamefont
  {Sokolov}}]{Roof16}%
  \BibitemOpen
  \bibfield  {author} {\bibinfo {author} {\bibfnamefont {S.~J.}\ \bibnamefont
  {Roof}}, \bibinfo {author} {\bibfnamefont {K.~J.}\ \bibnamefont {Kemp}},
  \bibinfo {author} {\bibfnamefont {M.~D.}\ \bibnamefont {Havey}}, \ and\
  \bibinfo {author} {\bibfnamefont {I.~M.}\ \bibnamefont {Sokolov}},\
  }\bibfield  {title} {\enquote {\bibinfo {title} {Observation of single-photon
  superradiance and the cooperative {L}amb shift in an extended sample of cold
  atoms},}\ }\href {\doibase 10.1103/PhysRevLett.117.073003} {\bibfield
  {journal} {\bibinfo  {journal} {Phys. Rev. Lett.}\ }\textbf {\bibinfo
  {volume} {117}},\ \bibinfo {pages} {073003} (\bibinfo {year}
  {2016})}\BibitemShut {NoStop}%
\bibitem [{\citenamefont {Corman}\ \emph {et~al.}(2017)\citenamefont {Corman},
  \citenamefont {Ville}, \citenamefont {Saint-Jalm}, \citenamefont
  {Aidelsburger}, \citenamefont {Bienaim\'e}, \citenamefont {Nascimb\`ene},
  \citenamefont {Dalibard},\ and\ \citenamefont {Beugnon}}]{COR17}%
  \BibitemOpen
  \bibfield  {author} {\bibinfo {author} {\bibfnamefont {L.}~\bibnamefont
  {Corman}}, \bibinfo {author} {\bibfnamefont {J.~L.}\ \bibnamefont {Ville}},
  \bibinfo {author} {\bibfnamefont {R.}~\bibnamefont {Saint-Jalm}}, \bibinfo
  {author} {\bibfnamefont {M.}~\bibnamefont {Aidelsburger}}, \bibinfo {author}
  {\bibfnamefont {T.}~\bibnamefont {Bienaim\'e}}, \bibinfo {author}
  {\bibfnamefont {S.}~\bibnamefont {Nascimb\`ene}}, \bibinfo {author}
  {\bibfnamefont {J.}~\bibnamefont {Dalibard}}, \ and\ \bibinfo {author}
  {\bibfnamefont {J.}~\bibnamefont {Beugnon}},\ }\bibfield  {title} {\enquote
  {\bibinfo {title} {Transmission of near-resonant light through a dense slab
  of cold atoms},}\ }\href {\doibase 10.1103/PhysRevA.96.053629} {\bibfield
  {journal} {\bibinfo  {journal} {Phys. Rev. A}\ }\textbf {\bibinfo {volume}
  {96}},\ \bibinfo {pages} {053629} (\bibinfo {year} {2017})}\BibitemShut
  {NoStop}%
\bibitem [{\citenamefont {Javanainen}\ \emph {et~al.}(1999)\citenamefont
  {Javanainen}, \citenamefont {Ruostekoski}, \citenamefont {Vestergaard},\ and\
  \citenamefont {Francis}}]{Javanainen1999a}%
  \BibitemOpen
  \bibfield  {author} {\bibinfo {author} {\bibfnamefont {Juha}\ \bibnamefont
  {Javanainen}}, \bibinfo {author} {\bibfnamefont {Janne}\ \bibnamefont
  {Ruostekoski}}, \bibinfo {author} {\bibfnamefont {Bjarne}\ \bibnamefont
  {Vestergaard}}, \ and\ \bibinfo {author} {\bibfnamefont {Matthew~R.}\
  \bibnamefont {Francis}},\ }\bibfield  {title} {\enquote {\bibinfo {title}
  {One-dimensional modelling of light propagation in dense and degenerate
  samples},}\ }\href@noop {} {\bibfield  {journal} {\bibinfo  {journal} {Phys.
  Rev. A}\ }\textbf {\bibinfo {volume} {59}},\ \bibinfo {pages} {649--666}
  (\bibinfo {year} {1999})}\BibitemShut {NoStop}%
\bibitem [{\citenamefont {Chomaz}\ \emph {et~al.}(2012)\citenamefont {Chomaz},
  \citenamefont {Corman}, \citenamefont {Yefsah}, \citenamefont {Desbuquois},\
  and\ \citenamefont {Dalibard}}]{CHO12}%
  \BibitemOpen
  \bibfield  {author} {\bibinfo {author} {\bibfnamefont {L}~\bibnamefont
  {Chomaz}}, \bibinfo {author} {\bibfnamefont {L}~\bibnamefont {Corman}},
  \bibinfo {author} {\bibfnamefont {T}~\bibnamefont {Yefsah}}, \bibinfo
  {author} {\bibfnamefont {R}~\bibnamefont {Desbuquois}}, \ and\ \bibinfo
  {author} {\bibfnamefont {J}~\bibnamefont {Dalibard}},\ }\bibfield  {title}
  {\enquote {\bibinfo {title} {Absorption imaging of a quasi-two-dimensional
  gas: a multiple scattering analysis},}\ }\href
  {http://stacks.iop.org/1367-2630/14/i=5/a=055001} {\bibfield  {journal}
  {\bibinfo  {journal} {New Journal of Physics}\ }\textbf {\bibinfo {volume}
  {14}},\ \bibinfo {pages} {055001} (\bibinfo {year} {2012})}\BibitemShut
  {NoStop}%
\bibitem [{\citenamefont {T.~Bienaim{\'{e}}}\ and\ \citenamefont
  {Kaiser}(2013)}]{Bienaime2013}%
  \BibitemOpen
  \bibfield  {author} {\bibinfo {author} {\bibfnamefont {N.~Piovella}\
  \bibnamefont {T.~Bienaim{\'{e}}}, \bibfnamefont {R.~Bachelard}}\ and\
  \bibinfo {author} {\bibfnamefont {R.}~\bibnamefont {Kaiser}},\ }\bibfield
  {title} {\enquote {\bibinfo {title} {Cooperativity in light scattering by
  cold atoms},}\ }\href@noop {} {\bibfield  {journal} {\bibinfo  {journal}
  {Fortschr. Phys.}\ }\textbf {\bibinfo {volume} {61}},\ \bibinfo {pages} {377}
  (\bibinfo {year} {2013})}\BibitemShut {NoStop}%
\bibitem [{\citenamefont {Javanainen}\ \emph {et~al.}(2014)\citenamefont
  {Javanainen}, \citenamefont {Ruostekoski}, \citenamefont {Li},\ and\
  \citenamefont {Yoo}}]{Javanainen2014a}%
  \BibitemOpen
  \bibfield  {author} {\bibinfo {author} {\bibfnamefont {Juha}\ \bibnamefont
  {Javanainen}}, \bibinfo {author} {\bibfnamefont {Janne}\ \bibnamefont
  {Ruostekoski}}, \bibinfo {author} {\bibfnamefont {Yi}~\bibnamefont {Li}}, \
  and\ \bibinfo {author} {\bibfnamefont {Sung-Mi}\ \bibnamefont {Yoo}},\
  }\bibfield  {title} {\enquote {\bibinfo {title} {Shifts of a resonance line
  in a dense atomic sample},}\ }\href {\doibase 10.1103/PhysRevLett.112.113603}
  {\bibfield  {journal} {\bibinfo  {journal} {Phys. Rev. Lett.}\ }\textbf
  {\bibinfo {volume} {112}},\ \bibinfo {pages} {113603} (\bibinfo {year}
  {2014})}\BibitemShut {NoStop}%
\bibitem [{\citenamefont {Sutherland}\ and\ \citenamefont
  {Robicheaux}(2016)}]{Sutherland1D}%
  \BibitemOpen
  \bibfield  {author} {\bibinfo {author} {\bibfnamefont {R.~T.}\ \bibnamefont
  {Sutherland}}\ and\ \bibinfo {author} {\bibfnamefont {F.}~\bibnamefont
  {Robicheaux}},\ }\bibfield  {title} {\enquote {\bibinfo {title} {Collective
  dipole-dipole interactions in an atomic array},}\ }\href {\doibase
  10.1103/PhysRevA.94.013847} {\bibfield  {journal} {\bibinfo  {journal} {Phys.
  Rev. A}\ }\textbf {\bibinfo {volume} {94}},\ \bibinfo {pages} {013847}
  (\bibinfo {year} {2016})}\BibitemShut {NoStop}%
\bibitem [{\citenamefont {Zhu}\ \emph {et~al.}(2016)\citenamefont {Zhu},
  \citenamefont {Cooper}, \citenamefont {Ye},\ and\ \citenamefont
  {Rey}}]{ZHU16}%
  \BibitemOpen
  \bibfield  {author} {\bibinfo {author} {\bibfnamefont {Bihui}\ \bibnamefont
  {Zhu}}, \bibinfo {author} {\bibfnamefont {John}\ \bibnamefont {Cooper}},
  \bibinfo {author} {\bibfnamefont {Jun}\ \bibnamefont {Ye}}, \ and\ \bibinfo
  {author} {\bibfnamefont {Ana~Maria}\ \bibnamefont {Rey}},\ }\bibfield
  {title} {\enquote {\bibinfo {title} {Light scattering from dense cold atomic
  media},}\ }\href {\doibase 10.1103/PhysRevA.94.023612} {\bibfield  {journal}
  {\bibinfo  {journal} {Phys. Rev. A}\ }\textbf {\bibinfo {volume} {94}},\
  \bibinfo {pages} {023612} (\bibinfo {year} {2016})}\BibitemShut {NoStop}%
\bibitem [{\citenamefont {Javanainen}\ \emph {et~al.}(2017)\citenamefont
  {Javanainen}, \citenamefont {Ruostekoski}, \citenamefont {Li},\ and\
  \citenamefont {Yoo}}]{JAV17}%
  \BibitemOpen
  \bibfield  {author} {\bibinfo {author} {\bibfnamefont {Juha}\ \bibnamefont
  {Javanainen}}, \bibinfo {author} {\bibfnamefont {Janne}\ \bibnamefont
  {Ruostekoski}}, \bibinfo {author} {\bibfnamefont {Yi}~\bibnamefont {Li}}, \
  and\ \bibinfo {author} {\bibfnamefont {Sung-Mi}\ \bibnamefont {Yoo}},\
  }\bibfield  {title} {\enquote {\bibinfo {title} {Exact electrodynamics versus
  standard optics for a slab of cold dense gas},}\ }\href {\doibase
  10.1103/PhysRevA.96.033835} {\bibfield  {journal} {\bibinfo  {journal} {Phys.
  Rev. A}\ }\textbf {\bibinfo {volume} {96}},\ \bibinfo {pages} {033835}
  (\bibinfo {year} {2017})}\BibitemShut {NoStop}%
\bibitem [{\citenamefont {Jackson}(1999)}]{Jackson}%
  \BibitemOpen
  \bibfield  {author} {\bibinfo {author} {\bibfnamefont {John~David}\
  \bibnamefont {Jackson}},\ }\href@noop {} {\emph {\bibinfo {title} {Classical
  Electrodynamics}}},\ \bibinfo {edition} {3rd}\ ed.\ (\bibinfo  {publisher}
  {Wiley, New York},\ \bibinfo {year} {1999})\BibitemShut {NoStop}%
\bibitem [{\citenamefont {Born}\ and\ \citenamefont {Wolf}(1999)}]{BOR99}%
  \BibitemOpen
  \bibfield  {author} {\bibinfo {author} {\bibfnamefont {Max}\ \bibnamefont
  {Born}}\ and\ \bibinfo {author} {\bibfnamefont {Emil}\ \bibnamefont {Wolf}},\
  }\href@noop {} {\emph {\bibinfo {title} {Principles of Optics}}},\ \bibinfo
  {edition} {7th}\ ed.\ (\bibinfo  {publisher} {Cambridge University Press,
  Cambridge, UK},\ \bibinfo {year} {1999})\BibitemShut {NoStop}%
\bibitem [{\citenamefont {Javanainen}\ and\ \citenamefont
  {Ruostekoski}(2016)}]{JavanainenMFT}%
  \BibitemOpen
  \bibfield  {author} {\bibinfo {author} {\bibfnamefont {Juha}\ \bibnamefont
  {Javanainen}}\ and\ \bibinfo {author} {\bibfnamefont {Janne}\ \bibnamefont
  {Ruostekoski}},\ }\bibfield  {title} {\enquote {\bibinfo {title} {Light
  propagation beyond the mean-field theory of standard optics},}\ }\href
  {\doibase 10.1364/OE.24.000993} {\bibfield  {journal} {\bibinfo  {journal}
  {Opt. Express}\ }\textbf {\bibinfo {volume} {24}},\ \bibinfo {pages}
  {993--1001} (\bibinfo {year} {2016})}\BibitemShut {NoStop}%
\bibitem [{\citenamefont {Jenkins}\ and\ \citenamefont
  {Ruostekoski}(2012)}]{Jenkins_2d_lat}%
  \BibitemOpen
  \bibfield  {author} {\bibinfo {author} {\bibfnamefont {Stewart~D.}\
  \bibnamefont {Jenkins}}\ and\ \bibinfo {author} {\bibfnamefont {Janne}\
  \bibnamefont {Ruostekoski}},\ }\bibfield  {title} {\enquote {\bibinfo {title}
  {Controlled manipulation of light by cooperative response of atoms in an
  optical lattice},}\ }\href {\doibase 10.1103/PhysRevA.86.031602} {\bibfield
  {journal} {\bibinfo  {journal} {Phys. Rev. A}\ }\textbf {\bibinfo {volume}
  {86}},\ \bibinfo {pages} {031602} (\bibinfo {year} {2012})}\BibitemShut
  {NoStop}%
\bibitem [{\citenamefont {Bettles}\ \emph {et~al.}(2016)\citenamefont
  {Bettles}, \citenamefont {Gardiner},\ and\ \citenamefont
  {Adams}}]{Bettles_prl16}%
  \BibitemOpen
  \bibfield  {author} {\bibinfo {author} {\bibfnamefont {Robert~J.}\
  \bibnamefont {Bettles}}, \bibinfo {author} {\bibfnamefont {Simon~A.}\
  \bibnamefont {Gardiner}}, \ and\ \bibinfo {author} {\bibfnamefont
  {Charles~S.}\ \bibnamefont {Adams}},\ }\bibfield  {title} {\enquote {\bibinfo
  {title} {Enhanced optical cross section via collective coupling of atomic
  dipoles in a 2{D} array},}\ }\href {\doibase 10.1103/PhysRevLett.116.103602}
  {\bibfield  {journal} {\bibinfo  {journal} {Phys. Rev. Lett.}\ }\textbf
  {\bibinfo {volume} {116}},\ \bibinfo {pages} {103602} (\bibinfo {year}
  {2016})}\BibitemShut {NoStop}%
\bibitem [{\citenamefont {Yoo}\ and\ \citenamefont {Paik}(2016)}]{YOO16}%
  \BibitemOpen
  \bibfield  {author} {\bibinfo {author} {\bibfnamefont {Sung-Mi}\ \bibnamefont
  {Yoo}}\ and\ \bibinfo {author} {\bibfnamefont {Sun~Mok}\ \bibnamefont
  {Paik}},\ }\bibfield  {title} {\enquote {\bibinfo {title} {Cooperative
  optical response of 2{D} dense lattices with strongly correlated dipoles},}\
  }\href {\doibase 10.1364/OE.24.002156} {\bibfield  {journal} {\bibinfo
  {journal} {Opt. Express}\ }\textbf {\bibinfo {volume} {24}},\ \bibinfo
  {pages} {2156--2165} (\bibinfo {year} {2016})}\BibitemShut {NoStop}%
\bibitem [{\citenamefont {Facchinetti}\ \emph {et~al.}(2016)\citenamefont
  {Facchinetti}, \citenamefont {Jenkins},\ and\ \citenamefont
  {Ruostekoski}}]{Facchinetti}%
  \BibitemOpen
  \bibfield  {author} {\bibinfo {author} {\bibfnamefont {G.}~\bibnamefont
  {Facchinetti}}, \bibinfo {author} {\bibfnamefont {S.~D.}\ \bibnamefont
  {Jenkins}}, \ and\ \bibinfo {author} {\bibfnamefont {J.}~\bibnamefont
  {Ruostekoski}},\ }\bibfield  {title} {\enquote {\bibinfo {title} {Storing
  light with subradiant correlations in arrays of atoms},}\ }\href {\doibase
  10.1103/PhysRevLett.117.243601} {\bibfield  {journal} {\bibinfo  {journal}
  {Phys. Rev. Lett.}\ }\textbf {\bibinfo {volume} {117}},\ \bibinfo {pages}
  {243601} (\bibinfo {year} {2016})}\BibitemShut {NoStop}%
\bibitem [{\citenamefont {Facchinetti}\ and\ \citenamefont
  {Ruostekoski}(2018)}]{FacchinettiLong}%
  \BibitemOpen
  \bibfield  {author} {\bibinfo {author} {\bibfnamefont {G.}~\bibnamefont
  {Facchinetti}}\ and\ \bibinfo {author} {\bibfnamefont {J.}~\bibnamefont
  {Ruostekoski}},\ }\bibfield  {title} {\enquote {\bibinfo {title} {Interaction
  of light with planar lattices of atoms: Reflection, transmission, and
  cooperative magnetometry},}\ }\href {\doibase 10.1103/PhysRevA.97.023833}
  {\bibfield  {journal} {\bibinfo  {journal} {Phys. Rev. A}\ }\textbf {\bibinfo
  {volume} {97}},\ \bibinfo {pages} {023833} (\bibinfo {year}
  {2018})}\BibitemShut {NoStop}%
\bibitem [{\citenamefont {Shahmoon}\ \emph {et~al.}(2017)\citenamefont
  {Shahmoon}, \citenamefont {Wild}, \citenamefont {Lukin},\ and\ \citenamefont
  {Yelin}}]{SHA17}%
  \BibitemOpen
  \bibfield  {author} {\bibinfo {author} {\bibfnamefont {Ephraim}\ \bibnamefont
  {Shahmoon}}, \bibinfo {author} {\bibfnamefont {Dominik~S.}\ \bibnamefont
  {Wild}}, \bibinfo {author} {\bibfnamefont {Mikhail~D.}\ \bibnamefont
  {Lukin}}, \ and\ \bibinfo {author} {\bibfnamefont {Susanne~F.}\ \bibnamefont
  {Yelin}},\ }\bibfield  {title} {\enquote {\bibinfo {title} {Cooperative
  resonances in light scattering from two-dimensional atomic arrays},}\ }\href
  {\doibase 10.1103/PhysRevLett.118.113601} {\bibfield  {journal} {\bibinfo
  {journal} {Phys. Rev. Lett.}\ }\textbf {\bibinfo {volume} {118}},\ \bibinfo
  {pages} {113601} (\bibinfo {year} {2017})}\BibitemShut {NoStop}%
\bibitem [{\citenamefont {Perczel}\ \emph
  {et~al.}(2017{\natexlab{a}})\citenamefont {Perczel}, \citenamefont
  {Borregaard}, \citenamefont {Chang}, \citenamefont {Pichler}, \citenamefont
  {Yelin}, \citenamefont {Zoller},\ and\ \citenamefont {Lukin}}]{PEROR17}%
  \BibitemOpen
  \bibfield  {author} {\bibinfo {author} {\bibfnamefont {J.}~\bibnamefont
  {Perczel}}, \bibinfo {author} {\bibfnamefont {J.}~\bibnamefont {Borregaard}},
  \bibinfo {author} {\bibfnamefont {D.~E.}\ \bibnamefont {Chang}}, \bibinfo
  {author} {\bibfnamefont {H.}~\bibnamefont {Pichler}}, \bibinfo {author}
  {\bibfnamefont {S.~F.}\ \bibnamefont {Yelin}}, \bibinfo {author}
  {\bibfnamefont {P.}~\bibnamefont {Zoller}}, \ and\ \bibinfo {author}
  {\bibfnamefont {M.~D.}\ \bibnamefont {Lukin}},\ }\bibfield  {title} {\enquote
  {\bibinfo {title} {Topological quantum optics in two-dimensional atomic
  arrays},}\ }\href {\doibase 10.1103/PhysRevLett.119.023603} {\bibfield
  {journal} {\bibinfo  {journal} {Phys. Rev. Lett.}\ }\textbf {\bibinfo
  {volume} {119}},\ \bibinfo {pages} {023603} (\bibinfo {year}
  {2017}{\natexlab{a}})}\BibitemShut {NoStop}%
\bibitem [{\citenamefont {Perczel}\ \emph
  {et~al.}(2017{\natexlab{b}})\citenamefont {Perczel}, \citenamefont
  {Borregaard}, \citenamefont {Chang}, \citenamefont {Pichler}, \citenamefont
  {Yelin}, \citenamefont {Zoller},\ and\ \citenamefont {Lukin}}]{PER17}%
  \BibitemOpen
  \bibfield  {author} {\bibinfo {author} {\bibfnamefont {J.}~\bibnamefont
  {Perczel}}, \bibinfo {author} {\bibfnamefont {J.}~\bibnamefont {Borregaard}},
  \bibinfo {author} {\bibfnamefont {D.~E.}\ \bibnamefont {Chang}}, \bibinfo
  {author} {\bibfnamefont {H.}~\bibnamefont {Pichler}}, \bibinfo {author}
  {\bibfnamefont {S.~F.}\ \bibnamefont {Yelin}}, \bibinfo {author}
  {\bibfnamefont {P.}~\bibnamefont {Zoller}}, \ and\ \bibinfo {author}
  {\bibfnamefont {M.~D.}\ \bibnamefont {Lukin}},\ }\bibfield  {title} {\enquote
  {\bibinfo {title} {Photonic band structure of two-dimensional atomic
  lattices},}\ }\href {\doibase 10.1103/PhysRevA.96.063801} {\bibfield
  {journal} {\bibinfo  {journal} {Phys. Rev. A}\ }\textbf {\bibinfo {volume}
  {96}},\ \bibinfo {pages} {063801} (\bibinfo {year}
  {2017}{\natexlab{b}})}\BibitemShut {NoStop}%
\bibitem [{\citenamefont {Asenjo-Garcia}\ \emph {et~al.}(2017)\citenamefont
  {Asenjo-Garcia}, \citenamefont {Moreno-Cardoner}, \citenamefont {Albrecht},
  \citenamefont {Kimble},\ and\ \citenamefont {Chang}}]{ASE17}%
  \BibitemOpen
  \bibfield  {author} {\bibinfo {author} {\bibfnamefont {A.}~\bibnamefont
  {Asenjo-Garcia}}, \bibinfo {author} {\bibfnamefont {M.}~\bibnamefont
  {Moreno-Cardoner}}, \bibinfo {author} {\bibfnamefont {A.}~\bibnamefont
  {Albrecht}}, \bibinfo {author} {\bibfnamefont {H.~J.}\ \bibnamefont
  {Kimble}}, \ and\ \bibinfo {author} {\bibfnamefont {D.~E.}\ \bibnamefont
  {Chang}},\ }\bibfield  {title} {\enquote {\bibinfo {title} {Exponential
  improvement in photon storage fidelities using subradiance and ``selective
  radiance'' in atomic arrays},}\ }\href {\doibase 10.1103/PhysRevX.7.031024}
  {\bibfield  {journal} {\bibinfo  {journal} {Phys. Rev. X}\ }\textbf {\bibinfo
  {volume} {7}},\ \bibinfo {pages} {031024} (\bibinfo {year}
  {2017})}\BibitemShut {NoStop}%
\bibitem [{\citenamefont {Bettles}\ \emph {et~al.}(2017)\citenamefont
  {Bettles}, \citenamefont {{Min\'a\ifmmode \check{r}\else \v{r}\fi{}}},
  \citenamefont {Adams}, \citenamefont {Lesanovsky},\ and\ \citenamefont
  {Olmos}}]{BET17}%
  \BibitemOpen
  \bibfield  {author} {\bibinfo {author} {\bibfnamefont {Robert~J.}\
  \bibnamefont {Bettles}}, \bibinfo {author} {\bibfnamefont {{Ji\ifmmode
  \check{r}\else \v{r}\fi{}\'{\i}}}\ \bibnamefont {{Min\'a\ifmmode
  \check{r}\else \v{r}\fi{}}}}, \bibinfo {author} {\bibfnamefont {Charles~S.}\
  \bibnamefont {Adams}}, \bibinfo {author} {\bibfnamefont {Igor}\ \bibnamefont
  {Lesanovsky}}, \ and\ \bibinfo {author} {\bibfnamefont {Beatriz}\
  \bibnamefont {Olmos}},\ }\bibfield  {title} {\enquote {\bibinfo {title}
  {Topological properties of a dense atomic lattice gas},}\ }\href {\doibase
  10.1103/PhysRevA.96.041603} {\bibfield  {journal} {\bibinfo  {journal} {Phys.
  Rev. A}\ }\textbf {\bibinfo {volume} {96}},\ \bibinfo {pages} {041603}
  (\bibinfo {year} {2017})}\BibitemShut {NoStop}%
\bibitem [{\citenamefont {Yoo}(2018)}]{YOO18}%
  \BibitemOpen
  \bibfield  {author} {\bibinfo {author} {\bibfnamefont {Sung-Mi}\ \bibnamefont
  {Yoo}},\ }\bibfield  {title} {\enquote {\bibinfo {title} {Strongly coupled
  cold atoms in bilayer dense lattices},}\ }\href {\doibase
  10.1088/1367-2630/aad614} {\bibfield  {journal} {\bibinfo  {journal} {New
  Journal of Physics}\ }\textbf {\bibinfo {volume} {20}},\ \bibinfo {pages}
  {083012} (\bibinfo {year} {2018})}\BibitemShut {NoStop}%
\bibitem [{\citenamefont {Jenkins}\ and\ \citenamefont
  {Ruostekoski}(2013)}]{CAIT}%
  \BibitemOpen
  \bibfield  {author} {\bibinfo {author} {\bibfnamefont {S.~D.}\ \bibnamefont
  {Jenkins}}\ and\ \bibinfo {author} {\bibfnamefont {J.}~\bibnamefont
  {Ruostekoski}},\ }\bibfield  {title} {\enquote {\bibinfo {title}
  {Metamaterial transparency induced by cooperative electromagnetic
  interactions},}\ }\href@noop {} {\bibfield  {journal} {\bibinfo  {journal}
  {Phys. Rev. Lett.}\ }\textbf {\bibinfo {volume} {111}},\ \bibinfo {pages}
  {147401} (\bibinfo {year} {2013})}\BibitemShut {NoStop}%
\bibitem [{\citenamefont {Jenkins}\ \emph {et~al.}(2017)\citenamefont
  {Jenkins}, \citenamefont {Ruostekoski}, \citenamefont {Papasimakis},
  \citenamefont {Savo},\ and\ \citenamefont {Zheludev}}]{JEN17}%
  \BibitemOpen
  \bibfield  {author} {\bibinfo {author} {\bibfnamefont {Stewart~D.}\
  \bibnamefont {Jenkins}}, \bibinfo {author} {\bibfnamefont {Janne}\
  \bibnamefont {Ruostekoski}}, \bibinfo {author} {\bibfnamefont {Nikitas}\
  \bibnamefont {Papasimakis}}, \bibinfo {author} {\bibfnamefont {Salvatore}\
  \bibnamefont {Savo}}, \ and\ \bibinfo {author} {\bibfnamefont {Nikolay~I.}\
  \bibnamefont {Zheludev}},\ }\bibfield  {title} {\enquote {\bibinfo {title}
  {Many-body subradiant excitations in metamaterial arrays: Experiment and
  theory},}\ }\href {\doibase 10.1103/PhysRevLett.119.053901} {\bibfield
  {journal} {\bibinfo  {journal} {Phys. Rev. Lett.}\ }\textbf {\bibinfo
  {volume} {119}},\ \bibinfo {pages} {053901} (\bibinfo {year}
  {2017})}\BibitemShut {NoStop}%
\bibitem [{\citenamefont {Deutsch}\ \emph {et~al.}(1995)\citenamefont
  {Deutsch}, \citenamefont {Spreeuw}, \citenamefont {Rolston},\ and\
  \citenamefont {Phillips}}]{DEU95}%
  \BibitemOpen
  \bibfield  {author} {\bibinfo {author} {\bibfnamefont {I.~H.}\ \bibnamefont
  {Deutsch}}, \bibinfo {author} {\bibfnamefont {R.~J.~C.}\ \bibnamefont
  {Spreeuw}}, \bibinfo {author} {\bibfnamefont {S.~L.}\ \bibnamefont
  {Rolston}}, \ and\ \bibinfo {author} {\bibfnamefont {W.~D.}\ \bibnamefont
  {Phillips}},\ }\bibfield  {title} {\enquote {\bibinfo {title} {Photonic band
  gaps in optical lattices},}\ }\href {\doibase 10.1103/PhysRevA.52.1394}
  {\bibfield  {journal} {\bibinfo  {journal} {Phys. Rev. A}\ }\textbf {\bibinfo
  {volume} {52}},\ \bibinfo {pages} {1394--1410} (\bibinfo {year}
  {1995})}\BibitemShut {NoStop}%
\bibitem [{\citenamefont {Le~Kien}\ \emph {et~al.}(2004)\citenamefont
  {Le~Kien}, \citenamefont {Balykin},\ and\ \citenamefont {Hakuta}}]{KIE04}%
  \BibitemOpen
  \bibfield  {author} {\bibinfo {author} {\bibfnamefont {Fam}\ \bibnamefont
  {Le~Kien}}, \bibinfo {author} {\bibfnamefont {V.~I.}\ \bibnamefont
  {Balykin}}, \ and\ \bibinfo {author} {\bibfnamefont {K.}~\bibnamefont
  {Hakuta}},\ }\bibfield  {title} {\enquote {\bibinfo {title} {Atom trap and
  waveguide using a two-color evanescent light field around a
  subwavelength-diameter optical fiber},}\ }\href {\doibase
  10.1103/PhysRevA.70.063403} {\bibfield  {journal} {\bibinfo  {journal} {Phys.
  Rev. A}\ }\textbf {\bibinfo {volume} {70}},\ \bibinfo {pages} {063403}
  (\bibinfo {year} {2004})}\BibitemShut {NoStop}%
\bibitem [{\citenamefont {Vetsch}\ \emph {et~al.}(2010)\citenamefont {Vetsch},
  \citenamefont {Reitz}, \citenamefont {Sagu\'e}, \citenamefont {Schmidt},
  \citenamefont {Dawkins},\ and\ \citenamefont
  {Rauschenbeutel}}]{Rauschenbeutel}%
  \BibitemOpen
  \bibfield  {author} {\bibinfo {author} {\bibfnamefont {E.}~\bibnamefont
  {Vetsch}}, \bibinfo {author} {\bibfnamefont {D.}~\bibnamefont {Reitz}},
  \bibinfo {author} {\bibfnamefont {G.}~\bibnamefont {Sagu\'e}}, \bibinfo
  {author} {\bibfnamefont {R.}~\bibnamefont {Schmidt}}, \bibinfo {author}
  {\bibfnamefont {S.~T.}\ \bibnamefont {Dawkins}}, \ and\ \bibinfo {author}
  {\bibfnamefont {A.}~\bibnamefont {Rauschenbeutel}},\ }\bibfield  {title}
  {\enquote {\bibinfo {title} {Optical interface created by laser-cooled atoms
  trapped in the evanescent field surrounding an optical nanofiber},}\ }\href
  {\doibase 10.1103/PhysRevLett.104.203603} {\bibfield  {journal} {\bibinfo
  {journal} {Phys. Rev. Lett.}\ }\textbf {\bibinfo {volume} {104}},\ \bibinfo
  {pages} {203603} (\bibinfo {year} {2010})}\BibitemShut {NoStop}%
\bibitem [{\citenamefont {Tiecke}\ \emph {et~al.}(2014)\citenamefont {Tiecke},
  \citenamefont {Thompson}, \citenamefont {de~Leon}, \citenamefont {Liu},
  \citenamefont {Vuleti{\'c}},\ and\ \citenamefont {Lukin}}]{TIE14}%
  \BibitemOpen
  \bibfield  {author} {\bibinfo {author} {\bibfnamefont {T.~G.}\ \bibnamefont
  {Tiecke}}, \bibinfo {author} {\bibfnamefont {J.~D.}\ \bibnamefont
  {Thompson}}, \bibinfo {author} {\bibfnamefont {N.~P.}\ \bibnamefont
  {de~Leon}}, \bibinfo {author} {\bibfnamefont {L.~R.}\ \bibnamefont {Liu}},
  \bibinfo {author} {\bibfnamefont {V.}~\bibnamefont {Vuleti{\'c}}}, \ and\
  \bibinfo {author} {\bibfnamefont {M.~D.}\ \bibnamefont {Lukin}},\ }\bibfield
  {title} {\enquote {\bibinfo {title} {Nanophotonic quantum phase switch with a
  single atom},}\ }\href {http://dx.doi.org/10.1038/nature13188} {\bibfield
  {journal} {\bibinfo  {journal} {Nature}\ }\textbf {\bibinfo {volume} {508}},\
  \bibinfo {pages} {241 EP --} (\bibinfo {year} {2014})}\BibitemShut {NoStop}%
\bibitem [{\citenamefont {Goban}\ \emph {et~al.}(2014)\citenamefont {Goban},
  \citenamefont {Hung}, \citenamefont {Yu}, \citenamefont {Hood}, \citenamefont
  {Muniz}, \citenamefont {Lee}, \citenamefont {Martin}, \citenamefont
  {McClung}, \citenamefont {Choi}, \citenamefont {Chang}, \citenamefont
  {Painter},\ and\ \citenamefont {Kimble}}]{kimblenaturecom}%
  \BibitemOpen
  \bibfield  {author} {\bibinfo {author} {\bibfnamefont {A.}~\bibnamefont
  {Goban}}, \bibinfo {author} {\bibfnamefont {C.~L.}\ \bibnamefont {Hung}},
  \bibinfo {author} {\bibfnamefont {S.~P.}\ \bibnamefont {Yu}}, \bibinfo
  {author} {\bibfnamefont {J.~D.}\ \bibnamefont {Hood}}, \bibinfo {author}
  {\bibfnamefont {J.~A.}\ \bibnamefont {Muniz}}, \bibinfo {author}
  {\bibfnamefont {J.~H.}\ \bibnamefont {Lee}}, \bibinfo {author} {\bibfnamefont
  {M.~J.}\ \bibnamefont {Martin}}, \bibinfo {author} {\bibfnamefont {A.~C.}\
  \bibnamefont {McClung}}, \bibinfo {author} {\bibfnamefont {K.~S.}\
  \bibnamefont {Choi}}, \bibinfo {author} {\bibfnamefont {D.~E.}\ \bibnamefont
  {Chang}}, \bibinfo {author} {\bibfnamefont {O.}~\bibnamefont {Painter}}, \
  and\ \bibinfo {author} {\bibfnamefont {H.~J.}\ \bibnamefont {Kimble}},\
  }\bibfield  {title} {\enquote {\bibinfo {title} {Atom--light interactions in
  photonic crystals},}\ }\href {http://dx.doi.org/10.1038/ncomms4808}
  {\bibfield  {journal} {\bibinfo  {journal} {Nat Commun}\ }\textbf {\bibinfo
  {volume} {5}},\ \bibinfo {pages} {3808} (\bibinfo {year} {2014})}\BibitemShut
  {NoStop}%
\bibitem [{\citenamefont {Douglas}\ \emph {et~al.}(2015)\citenamefont
  {Douglas}, \citenamefont {Habibian}, \citenamefont {Hung}, \citenamefont
  {Gorshkov}, \citenamefont {Kimble},\ and\ \citenamefont
  {Chang}}]{kimblemanybody}%
  \BibitemOpen
  \bibfield  {author} {\bibinfo {author} {\bibfnamefont {J.~S.}\ \bibnamefont
  {Douglas}}, \bibinfo {author} {\bibfnamefont {H.}~\bibnamefont {Habibian}},
  \bibinfo {author} {\bibfnamefont {C.~L.}\ \bibnamefont {Hung}}, \bibinfo
  {author} {\bibfnamefont {A.~V.}\ \bibnamefont {Gorshkov}}, \bibinfo {author}
  {\bibfnamefont {H.~J.}\ \bibnamefont {Kimble}}, \ and\ \bibinfo {author}
  {\bibfnamefont {D.~E.}\ \bibnamefont {Chang}},\ }\bibfield  {title} {\enquote
  {\bibinfo {title} {Quantum many-body models with cold atoms coupled to
  photonic crystals},}\ }\href {http://dx.doi.org/10.1038/nphoton.2015.57}
  {\bibfield  {journal} {\bibinfo  {journal} {Nat. Photon.}\ }\textbf {\bibinfo
  {volume} {9}},\ \bibinfo {pages} {326--331} (\bibinfo {year}
  {2015})}\BibitemShut {NoStop}%
\bibitem [{\citenamefont {Goban}\ \emph {et~al.}(2015)\citenamefont {Goban},
  \citenamefont {Hung}, \citenamefont {Hood}, \citenamefont {Yu}, \citenamefont
  {Muniz}, \citenamefont {Painter},\ and\ \citenamefont
  {Kimble}}]{kimblesuper}%
  \BibitemOpen
  \bibfield  {author} {\bibinfo {author} {\bibfnamefont {A.}~\bibnamefont
  {Goban}}, \bibinfo {author} {\bibfnamefont {C.-L.}\ \bibnamefont {Hung}},
  \bibinfo {author} {\bibfnamefont {J.~D.}\ \bibnamefont {Hood}}, \bibinfo
  {author} {\bibfnamefont {S.-P.}\ \bibnamefont {Yu}}, \bibinfo {author}
  {\bibfnamefont {J.~A.}\ \bibnamefont {Muniz}}, \bibinfo {author}
  {\bibfnamefont {O.}~\bibnamefont {Painter}}, \ and\ \bibinfo {author}
  {\bibfnamefont {H.~J.}\ \bibnamefont {Kimble}},\ }\bibfield  {title}
  {\enquote {\bibinfo {title} {Superradiance for atoms trapped along a photonic
  crystal waveguide},}\ }\href {\doibase 10.1103/PhysRevLett.115.063601}
  {\bibfield  {journal} {\bibinfo  {journal} {Phys. Rev. Lett.}\ }\textbf
  {\bibinfo {volume} {115}},\ \bibinfo {pages} {063601} (\bibinfo {year}
  {2015})}\BibitemShut {NoStop}%
\bibitem [{\citenamefont {S\o{}rensen}\ \emph {et~al.}(2016)\citenamefont
  {S\o{}rensen}, \citenamefont {B\'eguin}, \citenamefont {Kluge}, \citenamefont
  {Iakoupov}, \citenamefont {S\o{}rensen}, \citenamefont {M\"uller},
  \citenamefont {Polzik},\ and\ \citenamefont {Appel}}]{SOR16}%
  \BibitemOpen
  \bibfield  {author} {\bibinfo {author} {\bibfnamefont {H.~L.}\ \bibnamefont
  {S\o{}rensen}}, \bibinfo {author} {\bibfnamefont {J.-B.}\ \bibnamefont
  {B\'eguin}}, \bibinfo {author} {\bibfnamefont {K.~W.}\ \bibnamefont {Kluge}},
  \bibinfo {author} {\bibfnamefont {I.}~\bibnamefont {Iakoupov}}, \bibinfo
  {author} {\bibfnamefont {A.~S.}\ \bibnamefont {S\o{}rensen}}, \bibinfo
  {author} {\bibfnamefont {J.~H.}\ \bibnamefont {M\"uller}}, \bibinfo {author}
  {\bibfnamefont {E.~S.}\ \bibnamefont {Polzik}}, \ and\ \bibinfo {author}
  {\bibfnamefont {J.}~\bibnamefont {Appel}},\ }\bibfield  {title} {\enquote
  {\bibinfo {title} {Coherent backscattering of light off one-dimensional
  atomic strings},}\ }\href {\doibase 10.1103/PhysRevLett.117.133604}
  {\bibfield  {journal} {\bibinfo  {journal} {Phys. Rev. Lett.}\ }\textbf
  {\bibinfo {volume} {117}},\ \bibinfo {pages} {133604} (\bibinfo {year}
  {2016})}\BibitemShut {NoStop}%
\bibitem [{\citenamefont {Corzo}\ \emph {et~al.}(2016)\citenamefont {Corzo},
  \citenamefont {Gouraud}, \citenamefont {Chandra}, \citenamefont {Goban},
  \citenamefont {Sheremet}, \citenamefont {Kupriyanov},\ and\ \citenamefont
  {Laurat}}]{COR16}%
  \BibitemOpen
  \bibfield  {author} {\bibinfo {author} {\bibfnamefont {Neil~V.}\ \bibnamefont
  {Corzo}}, \bibinfo {author} {\bibfnamefont {Baptiste}\ \bibnamefont
  {Gouraud}}, \bibinfo {author} {\bibfnamefont {Aveek}\ \bibnamefont
  {Chandra}}, \bibinfo {author} {\bibfnamefont {Akihisa}\ \bibnamefont
  {Goban}}, \bibinfo {author} {\bibfnamefont {Alexandra~S.}\ \bibnamefont
  {Sheremet}}, \bibinfo {author} {\bibfnamefont {Dmitriy~V.}\ \bibnamefont
  {Kupriyanov}}, \ and\ \bibinfo {author} {\bibfnamefont {Julien}\ \bibnamefont
  {Laurat}},\ }\bibfield  {title} {\enquote {\bibinfo {title} {Large {B}ragg
  reflection from one-dimensional chains of trapped atoms near a nanoscale
  waveguide},}\ }\href {\doibase 10.1103/PhysRevLett.117.133603} {\bibfield
  {journal} {\bibinfo  {journal} {Phys. Rev. Lett.}\ }\textbf {\bibinfo
  {volume} {117}},\ \bibinfo {pages} {133603} (\bibinfo {year}
  {2016})}\BibitemShut {NoStop}%
\bibitem [{\citenamefont {Ruostekoski}\ and\ \citenamefont
  {Javanainen}(2016)}]{RUO16}%
  \BibitemOpen
  \bibfield  {author} {\bibinfo {author} {\bibfnamefont {Janne}\ \bibnamefont
  {Ruostekoski}}\ and\ \bibinfo {author} {\bibfnamefont {Juha}\ \bibnamefont
  {Javanainen}},\ }\bibfield  {title} {\enquote {\bibinfo {title} {Emergence of
  correlated optics in one-dimensional waveguides for classical and quantum
  atomic gases},}\ }\href {\doibase 10.1103/PhysRevLett.117.143602} {\bibfield
  {journal} {\bibinfo  {journal} {Phys. Rev. Lett.}\ }\textbf {\bibinfo
  {volume} {117}},\ \bibinfo {pages} {143602} (\bibinfo {year}
  {2016})}\BibitemShut {NoStop}%
\bibitem [{\citenamefont {Ruostekoski}\ and\ \citenamefont
  {Javanainen}(2017)}]{RUO17}%
  \BibitemOpen
  \bibfield  {author} {\bibinfo {author} {\bibfnamefont {Janne}\ \bibnamefont
  {Ruostekoski}}\ and\ \bibinfo {author} {\bibfnamefont {Juha}\ \bibnamefont
  {Javanainen}},\ }\bibfield  {title} {\enquote {\bibinfo {title} {Arrays of
  strongly coupled atoms in a one-dimensional waveguide},}\ }\href {\doibase
  10.1103/PhysRevA.96.033857} {\bibfield  {journal} {\bibinfo  {journal} {Phys.
  Rev. A}\ }\textbf {\bibinfo {volume} {96}},\ \bibinfo {pages} {033857}
  (\bibinfo {year} {2017})}\BibitemShut {NoStop}%
\bibitem [{\citenamefont {Chang}\ \emph {et~al.}(2004)\citenamefont {Chang},
  \citenamefont {Ye},\ and\ \citenamefont {Lukin}}]{CHA04}%
  \BibitemOpen
  \bibfield  {author} {\bibinfo {author} {\bibfnamefont {D.~E.}\ \bibnamefont
  {Chang}}, \bibinfo {author} {\bibfnamefont {Jun}\ \bibnamefont {Ye}}, \ and\
  \bibinfo {author} {\bibfnamefont {M.~D.}\ \bibnamefont {Lukin}},\ }\bibfield
  {title} {\enquote {\bibinfo {title} {Controlling dipole-dipole frequency
  shifts in a lattice-based optical atomic clock},}\ }\href {\doibase
  10.1103/PhysRevA.69.023810} {\bibfield  {journal} {\bibinfo  {journal} {Phys.
  Rev. A}\ }\textbf {\bibinfo {volume} {69}},\ \bibinfo {pages} {023810}
  (\bibinfo {year} {2004})}\BibitemShut {NoStop}%
\bibitem [{\citenamefont {Ruostekoski}\ and\ \citenamefont
  {Javanainen}(1997)}]{Ruostekoski1997a}%
  \BibitemOpen
  \bibfield  {author} {\bibinfo {author} {\bibfnamefont {Janne}\ \bibnamefont
  {Ruostekoski}}\ and\ \bibinfo {author} {\bibfnamefont {Juha}\ \bibnamefont
  {Javanainen}},\ }\bibfield  {title} {\enquote {\bibinfo {title} {Quantum
  field theory of cooperative atom response: Low light intensity},}\
  }\href@noop {} {\bibfield  {journal} {\bibinfo  {journal} {Phys. Rev. A}\
  }\textbf {\bibinfo {volume} {55}},\ \bibinfo {pages} {513--526} (\bibinfo
  {year} {1997})}\BibitemShut {NoStop}%
\bibitem [{\citenamefont {Press}\ \emph {et~al.}(2007)\citenamefont {Press},
  \citenamefont {Teukolski}, \citenamefont {Vetterling},\ and\ \citenamefont
  {Flannery}}]{NUMRES}%
  \BibitemOpen
  \bibfield  {author} {\bibinfo {author} {\bibfnamefont {W.~H.}\ \bibnamefont
  {Press}}, \bibinfo {author} {\bibfnamefont {S.~A.}\ \bibnamefont
  {Teukolski}}, \bibinfo {author} {\bibfnamefont {V.~A.}\ \bibnamefont
  {Vetterling}}, \ and\ \bibinfo {author} {\bibfnamefont {B.~P.}\ \bibnamefont
  {Flannery}},\ }\href@noop {} {\emph {\bibinfo {title} {Numerical Recipes: The
  art of scientific computing}}},\ \bibinfo {edition} {3rd}\ ed.\ (\bibinfo
  {publisher} {Cambridge University Press, NY},\ \bibinfo {year}
  {2007})\BibitemShut {NoStop}%
\bibitem [{Note1()}]{Note1}%
  \BibitemOpen
  \bibinfo {note} {There is a functionally equivalent figure in the
  Supplemental Material of Ref.~\cite {SHA17}}\BibitemShut {NoStop}%
\bibitem [{\citenamefont {Lewis}(1980)}]{LEW80}%
  \BibitemOpen
  \bibfield  {author} {\bibinfo {author} {\bibfnamefont {E.L.}\ \bibnamefont
  {Lewis}},\ }\bibfield  {title} {\enquote {\bibinfo {title} {Collisional
  relaxation of atomic excited states, line broadening and interatomic
  interactions},}\ }\href {\doibase
  http://dx.doi.org/10.1016/0370-1573(80)90056-3} {\bibfield  {journal}
  {\bibinfo  {journal} {Physics Reports}\ }\textbf {\bibinfo {volume} {58}},\
  \bibinfo {pages} {1 -- 71} (\bibinfo {year} {1980})}\BibitemShut {NoStop}%
\bibitem [{\citenamefont {Javanainen}(2017)}]{JAV17SA}%
  \BibitemOpen
  \bibfield  {author} {\bibinfo {author} {\bibfnamefont {Juha}\ \bibnamefont
  {Javanainen}},\ }\bibfield  {title} {\enquote {\bibinfo {title} {The
  {S}oftware {A}tom},}\ }\href {\doibase
  https://doi.org/10.1016/j.cpc.2016.09.017} {\bibfield  {journal} {\bibinfo
  {journal} {Computer Physics Communications}\ }\textbf {\bibinfo {volume}
  {212}},\ \bibinfo {pages} {1 -- 7} (\bibinfo {year} {2017})}\BibitemShut
  {NoStop}%
\bibitem [{\citenamefont {Prasad}\ and\ \citenamefont {Glauber}(2000)}]{PRA00}%
  \BibitemOpen
  \bibfield  {author} {\bibinfo {author} {\bibfnamefont {Sudhakar}\
  \bibnamefont {Prasad}}\ and\ \bibinfo {author} {\bibfnamefont {Roy~J.}\
  \bibnamefont {Glauber}},\ }\bibfield  {title} {\enquote {\bibinfo {title}
  {Polarium model: Coherent radiation by a resonant medium},}\ }\href {\doibase
  10.1103/PhysRevA.61.063814} {\bibfield  {journal} {\bibinfo  {journal} {Phys.
  Rev. A}\ }\textbf {\bibinfo {volume} {61}},\ \bibinfo {pages} {063814}
  (\bibinfo {year} {2000})}\BibitemShut {NoStop}%
\bibitem [{\citenamefont {Svidzinsky}\ \emph {et~al.}(2010)\citenamefont
  {Svidzinsky}, \citenamefont {Chang},\ and\ \citenamefont {Scully}}]{SVI10}%
  \BibitemOpen
  \bibfield  {author} {\bibinfo {author} {\bibfnamefont {Anatoly~A.}\
  \bibnamefont {Svidzinsky}}, \bibinfo {author} {\bibfnamefont {Jun-Tao}\
  \bibnamefont {Chang}}, \ and\ \bibinfo {author} {\bibfnamefont {Marlan~O.}\
  \bibnamefont {Scully}},\ }\bibfield  {title} {\enquote {\bibinfo {title}
  {Cooperative spontaneous emission of $n$ atoms: Many-body eigenstates, the
  effect of virtual {L}amb shift processes, and analogy with radiation of $n$
  classical oscillators},}\ }\href {\doibase 10.1103/PhysRevA.81.053821}
  {\bibfield  {journal} {\bibinfo  {journal} {Phys. Rev. A}\ }\textbf {\bibinfo
  {volume} {81}},\ \bibinfo {pages} {053821} (\bibinfo {year}
  {2010})}\BibitemShut {NoStop}%
\bibitem [{\citenamefont {Balik}\ \emph {et~al.}(2013)\citenamefont {Balik},
  \citenamefont {Win}, \citenamefont {Havey}, \citenamefont {Sokolov},\ and\
  \citenamefont {Kupriyanov}}]{Balik2013}%
  \BibitemOpen
  \bibfield  {author} {\bibinfo {author} {\bibfnamefont {S.}~\bibnamefont
  {Balik}}, \bibinfo {author} {\bibfnamefont {A.~L.}\ \bibnamefont {Win}},
  \bibinfo {author} {\bibfnamefont {M.~D.}\ \bibnamefont {Havey}}, \bibinfo
  {author} {\bibfnamefont {I.~M.}\ \bibnamefont {Sokolov}}, \ and\ \bibinfo
  {author} {\bibfnamefont {D.~V.}\ \bibnamefont {Kupriyanov}},\ }\bibfield
  {title} {\enquote {\bibinfo {title} {Near-resonance light scattering from a
  high-density ultracold atomic ${}^{87}${R}b gas},}\ }\href {\doibase
  10.1103/PhysRevA.87.053817} {\bibfield  {journal} {\bibinfo  {journal} {Phys.
  Rev. A}\ }\textbf {\bibinfo {volume} {87}},\ \bibinfo {pages} {053817}
  (\bibinfo {year} {2013})}\BibitemShut {NoStop}%
\end{thebibliography}
\end{document}